# Moiré Periodic and Quasiperiodic Crystals in Heterostructures of Twisted Bilayer Graphene and Hexagonal Boron Nitride

Xinyuan Lai,[1] Guohong Li,[1] Angela M. Coe,[1] Jedediah H. Pixley,[1,2,3] Kenji Watanabe,[4] Takashi Taniguchi,[4] Eva Y. Andrei[1*]

[1]Department of Physics and Astronomy, Rutgers University, 136 Frelinghuysen Road, Piscataway, NJ 08854, USA

[2]Center for Computational Quantum Physics, Flatiron Institute, 162 Fifth Avenue, New York, NY 10010, USA

[3]Center for Materials Theory, Rutgers University, 136 Frelinghuysen Road, Piscataway, NJ 08854, USA

[4]Research Center for Functional Materials, National Institute for Materials Science, 1-1 Namiki, Tsukuba 305-0044, Japan

∗To whom correspondence should be addressed, E-mail: eandrei@physics.rutgers.edu

**Abstract:**
Stacking two atomic crystals with a twist between their crystal axes produces moiré potentials that modify the electronic properties. Here we show that double moiré potentials generated by superposing three atomic crystals create a new class of tunable quasiperiodic structures that alter the symmetry and spatial distribution of the electronic wavefunctions. By using scanning tunneling microscopy and spectroscopy to study twisted bilayer graphene on hexagonal boron nitride (hBN), we unveil a moiré phase diagram defined by the lattice constants of the two moiré lattices (graphene-on-graphene and graphene-on-hBN), comprising both commensurate periodic and incommensurate quasiperiodic crystals. Remarkably, the 1:1 commensurate crystal, which should theoretically exist at only one point on this phase diagram, is observed over a wide range, demonstrating an unexpected self-alignment mechanism. The incommensurate crystals include quasicrystals, which are quasiperiodic and feature a Bravais-forbidden dodecagonal symmetry, and intercrystals, which are also quasiperiodic but lack forbidden symmetries. This rich variety of tunable double moiré structures offers a synthetic platform for exploring the unique electronic properties of quasiperiodic crystals, which are rarely found in nature.

**Main text:**

Moiré potentials generated by stacking two atomic crystals can profoundly affect the electronic properties[1-9]. Superposing three identical atomic crystals such as trilayer graphene, gives rise to new moiré structures and flat bands[10] leading to emergent correlated phases[11-14]. Here we show that superposing three non-identical layered crystals with a twist between their crystal axes generates an even richer variety of fascinating structures and electronic properties.

In double-layer graphene (GG), a twist of $\theta_{GG} \approx 1°$ between the layers - known as the magic angle - produces a nearly flat band with non-trivial topology. Breaking either the time-reversal symmetry with a magnetic field or the sublattice symmetry ($C_{2z}$) with a staggered potential reveals Chern insulating states at integer fillings of the moiré unit cell [15-18]. In contrast to the readily observed Chern insulators in an external magnetic field [19-21], those induced when the $C_{2z}$ symmetry is broken by the staggered potential created when aligning twisted bilayer graphene with an hBN crystal are rare, their hallmark anomalous quantum Hall effect (AQHE) having been reported in only two samples[15,16]. A possible reason for this scarcity is the lattice mismatch between graphene and hBN, $a_G = 0.2461 nm$, $a_{BN} = 0.2504 nm$, as a result of which 1:1 alignment between magic angle GG and graphene-on-hBN (GBN), can only occur for three distinct values of the twist angle: $\theta_{GBN} = \pm 0.55°, 0°$ [22-25]. For $\theta_{GBN} = +0.55°$ the three layers are rotated in the same sense corresponding to helical stacking, while for $\theta_{GBN} = -0.55°$ adjacent layers are rotated in opposite sense, corresponding to mirror symmetric or anti-helical stacking. While techniques exist for precisely aligning two graphene layers cut from the same flake[26], aligning graphene and hBN over a several-micron sample area is not yet possible. This raises the question of what enabled the observation of the AQHE in magic angle GG aligned with hBN (GG/GBN).

Using low-temperature scanning-tunneling-microscopy (STM) and spectroscopy (STS)[27,28] (Methods) to study the double-moiré structures formed in GG/GBN, we have identified periodic and quasiperiodic crystals and characterized their electronic properties through STM topography and local density of states (LDoS) estimated from dI/dV measurements. The samples, prepared by depositing twisted bilayer graphene on a bulk hBN (~30nm) substrate (Methods), span a range of twist angles $\theta_{GBN} < 1.7°$ and $\theta_{GG} \approx 0.5 - 3.5°$ corresponding to GBN moiré lattice constants $L_{GBN}$ ~ 4-15 nm and GG lattice constants $L_{GG}$ ~2-40 nm. Surprisingly 1:1 commensurate crystals, where $L_{GBN} = L_{GG}$, are observed over a much wider range of twist-angle pairs than expected for rigid lattices, indicating the presence of an underlying self-alignment mechanism. Analysis of the double moiré structures reveals a phase diagram (Fig. 3) spanned by the two moiré lattice constants, comprised of lines of commensurate moiré crystals embedded in a background of incommensurate quasiperiodic crystals. This rich phase diagram provides a road map for generating 2D moiré periodic and quasiperiodic crystals by tuning the twist angles.

The 1:1 commensurate crystal are periodic lattices spanned by two independent lattice vectors (rank 2). They obey Bloch's theorem leading to well-defined crystal momentum, and they display only 2, 3 and 6 fold rotational symmetries. Conversely, the incommensurate lattices are spanned by more than 2 lattice vectors (rank > 2), they have no translational symmetry, and do not obey the Bloch theorem[29-34]. In the absence of translational symmetry long range order can still exist leading to sharp Bragg peaks, but the structure is quasiperiodic. Quasiperiodic crystals come in two varieties: moiré quasicrystals (MQCs) and moiré intercrystals (MICs). MQCs are distinguished by their Bravais-forbidden dodecagonal rotational symmetry[30,31], while MICs exhibit only the permissible rotational symmetries. The absence of forbidden symmetries in MICs positions them between crystals and quasicrystals in terms of their symmetry properties, which is why we refer to them as "intercrystals".

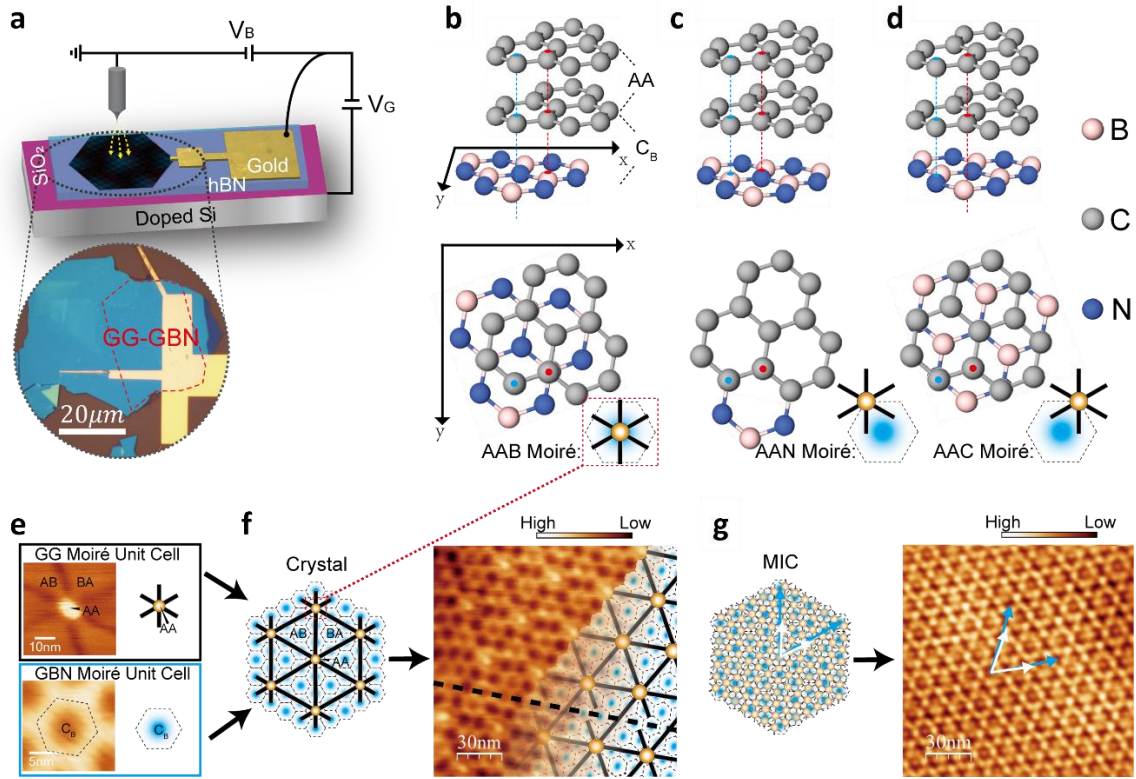

Fig. 1 **Imaging GG and GBN double moiré pattern and the preferred local stacking order. a,** Schematic drawing of the experimental setup. Sample bias $V_B$ is applied to the gold electrode. Sample doping is tuned by the gate voltage, $V_G$, applied to the silicon substrate, and the moiré pattern is imaged with the scanning tip. Inside the circle is an optical micrograph of the device. **b-d,** Schematic atomic configuration of different stacking orders: **b,** AAB, **c,** AAN, **d,** AAC. The top and bottom panels correspond to 3D and top views of the model respectively. Boron, carbon, and nitrogen atoms are labeled by pink, grey and blue circles respectively. Red and blue spots mark the graphene A and B sublattices respectively. The lower right insets schematically represent the different stackings as viewed in STM topography. **e,** STM topography and schematic drawing of GG and GBN moiré patterns. The GG moiré unit cell shown in the top left panel and schematically to its right, consists of a bright central AA region surrounded by six alternating less bright Bernal stacked (AB/BA) regions. The smaller area of AA compared to AB/BA regions reflects the lattice relaxation towards the energetically favored Bernal stacking. The bottom left panel and the schematic inset on the right show the STM topography of a single GBN moiré cell consisting of a central dark spot. The schematic in the right panel illustrates the AAB stacking. **f,** Left panel: STM topography showing an example of a moiré crystal in AAB stacking configuration ( $L_{GG}$=33.6 nm; $L_{GBN}$=11.2 nm). Right corner of the panel shows superposed schematic patterns for $L_{GG}$=3$L_{GBN}$. Dashed line represents a domain boundary from the relaxation of double moiré patterns. **g,** Right panel: STM topography showing an example of the moiré intercrystal (right panel; $L_{GG}$=7.1 nm; $L_{GBN}$=11.6 nm) and its schematic representation (left panel). The GG ($\vec{a}_{GG}$) and GBN ($\vec{a}_{GBN}$) lattice vectors are marked by white and blue arrows, respectively. Tunneling parameters: bias voltage $V_B$ = -500mV, gate voltage $V_G$= 0V, tunneling current I = 20pA.

Fig. 1 illustrates the three local stacking orders between the AA sites of GG and the underlying hBN lattice. These are labeled by AAB, AAN, or AAC corresponding to AA aligned with either the B atom, the N atom, or the center of the hBN unit cell, as shown in Fig. 1b, 1c, and 1d, respectively. The stacking order is revealed in STM topography by imaging the two moiré patterns simultaneously, as illustrated in the Fig. 1e-g. The GG pattern is identified by the bright spots produced by the AA stacking, while the GBN pattern displays dark spots, indicative of $C_B$ a Carbon atom aligned with a Boron atom stacking as illustrated in Fig. 1e. Thus, the 1:1 commensurate crystal, where all GBN sites coincide with AA sites will present a triangular lattice of bright spots and no dark spots, providing a direct image of the GG lattice, as shown in Fig. 2a. If both bright and dark spots appear in the topography, the structure is identified from the two-lattice constants, where bright spots and dark spots span the GG and GBN lattices, respectively. The examples in Fig.1f and Fig. 1g, show a 1:3 commensurate crystal, and a rank-4 MIC with its 4 incommensurate lattice vectors marked by arrows. Our study, comprised of dozens of twist angle pairs, reveals large areas of contiguous AAB stacking, indicating commensurability, even for samples whose twist angle deviates substantially from the rigid lattice commensuration (RLC) condition. This suggests the presence of a relaxation mechanism causing the GG and GBN moiré patterns to self-align. Indeed, according to first principles calculations [35,36], the AA sites of GG have the highest onsite energy while the $C_B$ sites of GBN the lowest energy, so that the AAB configuration is likely to be energetically favored. This conclusion is supported by comparing the topography obtained from forward and backward STM scans in regions with different stacking orders (Fig. S3). The topography of AAB domains is identical in forward and backward scans, while it changes with scan direction for other stacking orders. A relative shift of the local stacking triggered by the interaction between the sample and moving STM tip for non-AAB stacked sites indicates instability.

In the small angle approximation valid for the samples in this study, the two moiré lattice constants are given by:

$L_{GG} = \frac{a_G}{2\sin\left(\frac{\theta_{GG}}{2}\right)} \approx \frac{a_G}{\theta_{GG}}$ and $L_{GBN} = \frac{(1+\delta)a_G}{\sqrt{2(1+\delta)(1-\cos(\theta_{GBN}))+\delta^2}} \approx \frac{a_{BN}}{\sqrt{(\theta_{GBN})^2+\delta^2}}$, where $\delta = \frac{a_{BN}}{a_G} - 1 = 0.017$ is the lattice mismatch. Experimentally we obtain their values from the fast Fourier transform (FFT) of the topography by using the average magnitude of the six Bragg peak pairs, $|\vec{K}_{GG}|$ and $|\vec{K}_{GBN}|$ (in radians per unit length) to calculate $L_{GG} = 2/(\sqrt{3}|\vec{K}_{GG}|)$ and $L_{GBN} = 2/(\sqrt{3}|\vec{K}_{GBN}|)$.

The 1:1 commensurate crystal share a common moiré wavelength labeled $L_M = L_{GBN} = L_{GG}$. Fig. 2a-c show the topography of three such lattices, with $L_M = 12.8\ nm, 10.6\ nm$, and $10.0\ nm$ respectively. As expected for 1:1 commensurate lattices the topography reveals a triangular lattice of bright AAB stacking points and no visible dark ($C_B$) stacking points. Their FFTs display 6 primary Bragg peaks produced by the perfectly aligned GG and GBN moiré patterns which are accompanied by 12 second-order peaks. Structurally these are rank-2 crystals with 6-fold rotational symmetry. For rigid lattices the 1:1 RLC moiré wavelength is $L_M = 12.8\ nm$ (see

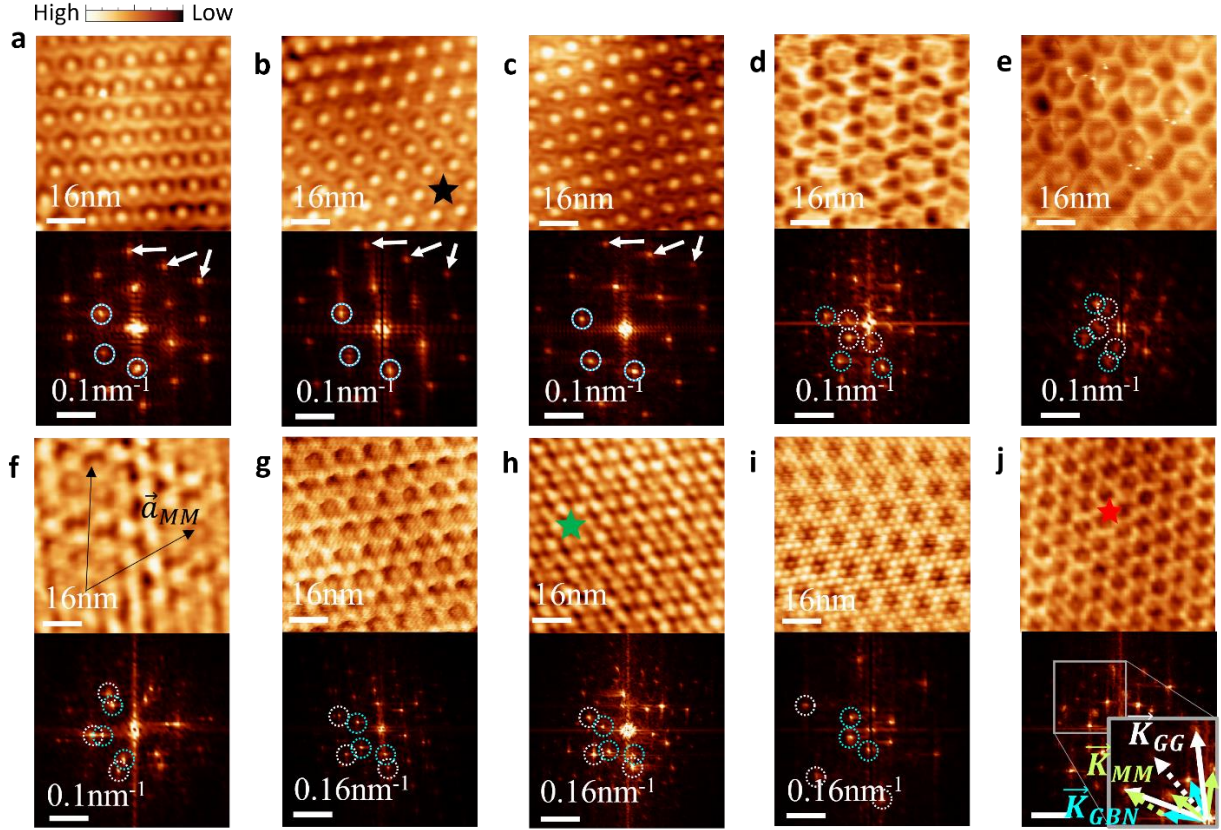

Fig. 2. **Double moiré patterns.** STM topography and the corresponding FFT are shown in the top and bottom panels, respectively. **a-c** 1:1 commensurate moiré crystals: with $L_M = 12.8\ nm$ ($\theta_{GG} = 1.10°; \theta_{GBN} = 0.55°$), $L_M = 10.6\ nm$ ($\theta_{GG} = 1.33°; \theta_{GBN} = 0.93°$) and $L_M = 10.0\ nm$ ($\theta_{GG} = 1.41°; \theta_{GBN} = 1.04°$), respectively. White arrows point to second-order Bragg peaks. **d-e,** Moiré intercrystals (MIC) with $L_{GG} \approx 18\ nm; L_{GBN} \approx 13\ nm$ and $L_{GG} \approx 22\ nm; L_{GBN} \approx 13\ nm$, respectively. In both cases $L_{GG} > L_{GBN}$ and the Bragg peaks are broadened reflecting a higher susceptibility to local perturbations as discussed in the text. **f,** Moiré intercrystal (MIC) with $L_{GG} = 10.7\ nm; L_{GBN} = 14.5\ nm$. ($\theta_{GG} = 1.32°; \theta_{GBN} \approx 0°$) showing an obvious moiré-of-moiré pattern whose lattice vectors $\vec{a}_{MM}$ are marked with black arrows. **g-i** Moiré Intercrystals (MIC): $L_{GG} = 7.8nm; L_{GBN} = 11.3\ nm$ ($\theta_{GG} = 1.8°; \theta_{GBN} = 0.8°$), $L_{GG} = 7.1nm; L_{GBN} = 10.9\ nm$, ($\theta_{GG} = 1.98°; \theta_{GBN} = 1.02°$) and $L_{GG} = 4.0\ nm; L_{GBN} = 13.3\ nm$ ($\theta_{GG} = 3.52°; \theta_{GBN} = 0.45°$) respectively. Both GG and GBN periods are visible and the corresponding reciprocal lattice vectors $\vec{K}_{GG}$ and $\vec{K}_{GBN}$ are marked in the FFT with white and blue circles, respectively. **j,** Moiré quasircrystal (MQC): $L_{GG} = 4.55\ nm; L_{GBN} = 9.42\ nm$. ($\theta_{GG} = 3.10°; \theta_{GBN} = 1.16°$) and its FFT. Inset: Zoom-in to the grey rectangle of the FFTs. The GG, GBN, and moiré-of-moiré wavevectors, $\vec{K}_{GG}, \vec{K}_{GBN}$ and $\vec{K}_{MM}$, are marked with white, blue and green arrows. The scale bar is $0.16\ nm^{-1}$. Tunneling parameters: bias voltage $V_B$ = -500mV; gate voltage $V_G$ = 0V; I = 20pA.

Fig. 3 and SI). We stress that, in the absence of strain the RLC for the 1:1 commensurate lattice has a unique wavelength of 12.8 nm. We note that only the crystal in Fig. 2a coincides with the RLC wavelength, implying that the lattice is unstrained. Conversely, all 1:1 moiré crystals with $L_M \neq 12.8nm$ must be strained to achieve commensuration, thus providing a much larger range of effectively commensurate structures through relaxation.

In addition to the 1:1 commensurate crystal, we observe quasiperiodic crystals identified as MICs according to the criteria: (i) FFT of topography maps display well-defined Bragg peaks, reflecting the long-range order. (ii) the lattice is spanned by more than two linearly independent vectors giving rise to quasi-periodicity. (iii) The lattice exhibits only the allowed rotational symmetries for this system: 2,3, or 6-fold.

Examples of MICs are shown in Figs 2d-i. Quasi-periodicity arises when the ratio between a lattice vector and its projection along another lattice vector, $p \equiv L_{GG} \cos\beta / L_{GBN}$, is irrational. Here $\beta$ is the angle between the vectors. For the example in Fig. 2i where $L_{GG} = 4.0 \ nm$, $L_{GBN} = 13.3\ nm$ and $\beta = 4.6°$, we have p= 0.3005336... in contrast to the moiré crystals where the two pairs of lattice vectors overlap.

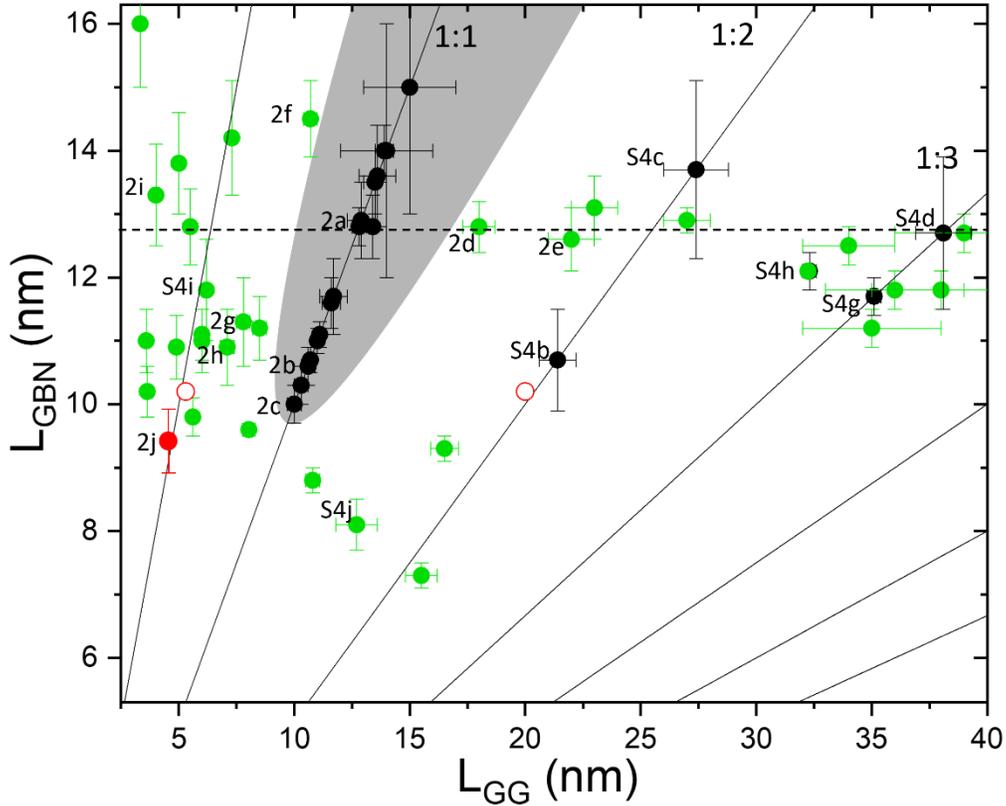

**Figure 3.** $L_{GBN}$ versus $L_{GBG}$ phase diagram classifying the double moiré patterns into moiré crystals (black dots), moiré intercrystals (MIC green dots) and moiré quasicrystals (MQC red). Solid black lines indicating commensurate lattices $L_{GG} = \frac{n}{m} L_{GBN}; n, m \in \mathbb{Z}^+$ are labeled with their respective n:m ratio. The horizontal dashed line at $L_{GBN} = 12.8 \ nm$ marks the condition for which rigid lattice commensuration (RLC) may occur, when intersecting with the solid black lines where $L_{GG} = \frac{n}{m} 12.8 \ nm$. Topographies in Fig. 2 and Fig. S4 are labeled by the figure number. The observed crystals coincide with the solid lines, but many are quite far from the RLC points. The expected positions of the two unstrained dodecagonal MQC are marked by open red stars. The slightly strained dodecagonal MQC observed experimentally (Fig. 2i) is marked by the solid red spot. The error margin is estimated by the resolution of FFT (SI). Bragg peaks corresponding to $\vec{K}_{MM}$ in **j** are amplified by a factor of 3 for clarity. The gray shadow area around the 1:1 commensuration line represents the basin of attraction – all lattices with moiré wavelength within this region relax to the 1:1 commensuration line.

Interference between the GBN and GG moiré patterns may produce secondary moiré-of-moiré super-patterns of period $L_{MM}$ arising from linear combinations of their reciprocal lattice vectors. As a higher order interference effect the moiré-of-moiré patterns and corresponding Bragg peaks in the FFT are weaker than the primary ones and not always well resolved. In Fig 2f, we show an example where the moiré-of-moiré patterns are clearly seen in the real space topography ($\vec{a}_{MM}$ lattice vectors shown) but the associated Bragg peaks in FFT are not well resolved.

The collected data sets are classified in the $[L_{GBN}, L_{GG}]$ phase diagram shown in Fig. 3. The solid black lines corresponding to commensurate moiré lattices are labeled by the commensuration ratio n:m. The intersection of the horizontal dashed line ($L_{GBN} = 12.8 nm$) with the n:m commensuration lines mark the RLC conditions. We note that many of the commensurate data points (solid black dots) deviate substantially, by as much as 20% from the RLC condition $L_{GBN}=L_{GG} = 12.8 nm$ providing clear evidence of self-alignment. The region devoid of data points flanking the 1:1 line, represents the self-alignment basin of attraction (gray area in Fig. 3) which sets the boundary between commensurate and incommensurate structures. Other commensurate crystals are observed on the 1:2, 1:3, 1:4, and 1:5 lines (Fig. S4). These crystals are forbidden for rigid lattices suggesting a self-alignment process involving in-plane strain of the atomic layers that homogeneously change the lattice mismatch between graphene and hBN. In contrast to the previous studies of the relaxation within a single moiré unit cell[37,38], global self-alignment requires homogeneous straining of the atomic layers over many moiré cells.

To understand the self-alignment mechanism, we use an atomic scale semi-classical model based on the competition between the strain energy cost and the Van der Waals energy gain from alignment in the preferred AAB stacking order. The model assumes that the graphene monolayers relax while hBN is rigid. Plotting the Brillouin zone of the 1:1 commensurate RLC condition for helical stacking ($\theta_{GG} = 1.1°, \theta_{GBN} = +0.55°$) in Fig. 3a we note that the twist-angle between the reciprocal-lattice vectors of the GG and GBN rigid lattices ($\vec{K}_{GG0}$ and $\vec{K}_{GBN0}$) is $\varphi = 120°$ (Fig. 3a inset), while the anti-helical RLC ($\theta_{GG} = 1.1°$, $\theta_{GBN} = -0.55°$) gives $\varphi = 60°$ (Fig. S8). In general, misaligned lattices will have $|\theta_{GBN}| \neq \frac{\theta_{GG}}{2}$ and $\varphi \neq 120°$ (or 60°). Global commensuration would thus require rescaling of the lattice vectors so that $K_{GG} = K_{GBN}$, and $\varphi = 120°$ (or 60°) [39].

We illustrate the self-alignment mechanism for two cases, symmetric and asymmetric. In the symmetric case with $|\theta_{GBN}| = \frac{\theta_{GG}}{2}$ but $K_{GG} \neq K_{GBN}$ (Fig. 4b, 4c top panels), both top and bottom graphene layers stretch or contract equally so that for both layers $\Delta K_G = \frac{\sqrt{3}}{2}(K_{GG0} - K_{GG})$, resulting in a symmetric atomic-strain of $\varepsilon_s = \frac{\Delta K_G}{K_{G0}} = \frac{\sqrt{3}}{2}\frac{a_G}{L_M}\left(\frac{\Delta L_M}{L_{M0}}\right)$. In the asymmetric case, corresponding to heterostrain, $|\theta_{GBN}| \neq \frac{\theta_{GG}}{2}$ and $K_{GG} \neq K_{GBN}$ (Fig. 4b, 4c bottom panels), self-alignment requires adding an asymmetric contribution to the symmetric $\Delta K_G$, different for the

top and bottom layers, $\Delta K_{G-top} \neq \Delta K_{G-bot}$, so that $\varepsilon_{bot} = \frac{\Delta K_G + \Delta K_{G-bot}}{K_{G0}}$ and $\varepsilon_{top} = \frac{\Delta K_G + \Delta K_{G-top}}{K_{G0}}$ (Fig. 4c bottom panel). The calculated elastic-energy cost per moiré cell for aligning the two lattices in this model, $E_{el}$, is shown for $\varphi = 120°$ in Figs. 4d, 4e. Plotting the dependence of $E_{el}$ on $L_M$ for both 60° and 120° cases (Fig. 3f) it is clear that the latter is energetically favored. By comparing the Van der Waals energy gain of AAB stacking over the area of AA sites with the elastic energy cost of straining [40] (Methods), we obtain an estimated upper bound of $E_{el}$ (pink line in Fig. 4f) for completely relaxed lattices. This gives the range of $L_M$ values, [10 nm, 16 nm] for which the *120°* global self-alignment is energetically favored, with the exception of $L_M \sim L_{M0}$ where *60°* and *120°* are equally likely. Projecting the measured $L_M$ values of the 1:1 aligned lattices in Fig. 3 onto the calculated curves, we find that this boundary closely matches the experimental basin of attraction (gray area in Fig 3).

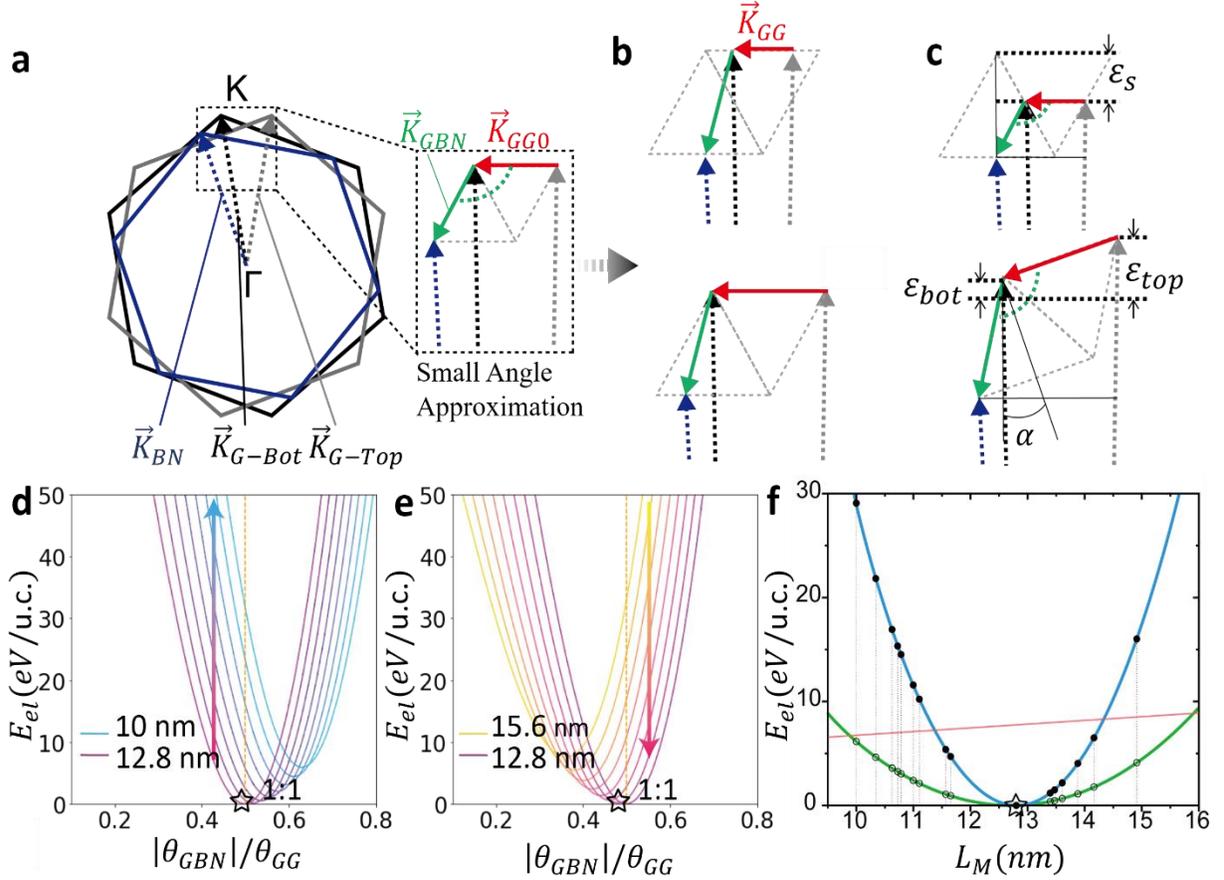

Fig. 4. **Mechanism of moiré self-alignment. a,** Schematic drawings of reciprocal lattice vectors of top and bottom graphene layers ($\vec{K}_{G-Top}$, $\vec{K}_{G-Bot}$) and hBN ($\vec{K}_{hBN}$). The zoomed-in inset shows that the 1:1 at $120^0$ commensuration occurs when the moiré reciprocal lattice vectors of GG and GBN, $\vec{K}_{GG}$ (red arrow) and $\vec{K}_{GBN}$ green arrow respectively, form a rhombus with inner angles $\varphi=120°$. Note that owing to the small angle approximation the reciprocal lattice vectors appear to be parallel to each in the zoomed in view. **b,** Schematic drawing of incommensurate cases with $\theta_{GBN} \neq 1.1°$ (top panel) and $\theta_{GBN} \neq |\frac{1}{2}\theta_{GG}|$ (bottom panel). **c,** Schematic drawings of self-alignment with symmetric (top) strain $\varepsilon_s$ or asymmetric (bottom) strain defined different for the bottom and top layer as $\varepsilon_{bot}$ and $\varepsilon_{top}$ as described in the text and ref. The tilt angle, $\alpha$ is defined by $\tan \alpha = \left(\theta_{GBN} - \frac{1}{2}\theta_{GG}\right) * \frac{|\vec{K}_{GBN}|}{\frac{\sqrt{3}}{2}|\vec{K}_{GG}|}$. **d-e,** Calculated elastic energy of alignment versus twist-angle ratio for 1:1 commensuration at *120°* for several moiré wavelength $L_M$ <12.8nm and $L_M$ >12.8nm. **f,** Calculated lowest elastic energy for self-aligned commensuration (blue curve for *60°*; green curve for *120°*) as a function of $L_M$. The experimentally observed $L_M$ values are estimated assuming they are either *60°* or *120°* commensuration and plotted with solid and empty circles, respectively. The pink curve marks the estimated energy limit for self-alignment (see SI for details). Thus, the *120°* alignment is energetically favored for most data points except for $L_M \sim L_{M0}$ where *60°* and *120°* are equally likely. Note that owing to the small angle approximation the reciprocal lattice vectors appear to be parallel to each other in the zoomed in view.

Most data points that are outside the 1:1 basin of attraction represent MICs. Those to the left of the 1:1 line ($L_{GG} < L_{GBN}$) feature well defined GG and GBN periods as evidenced by the sharp peaks in the FFT of the topography (Figs. 2f-i). For MICs to the right of the 1:1 line ($L_{GG} > L_{GBN}$) where the large GG moiré cells render the system more susceptible to perturbations such as twist angle disorder or strain, the local AAB stacking order is preferred over long-range order (Figs 2d-e) . The local AAB stacking shifts the AA positions to the nearest $C_B$ site, forming locally aligned structures with $L_{GG} = n * L_{GBN}; n \in \mathbb{Z}^+$. In this regime the competition between extrinsic disorder and self-alignment disrupts the long-range order. This may produce glassy structures (Fig. S7) leading to smeared GG Bragg peaks.

The GG/GBN quasiperiodic moiré crystals are highly tunable and under certain conditions they may acquire a Bravais-forbidden rotational symmetry resulting in a quasicrystal structure[29-31]. Since the double-moiré lattices are generated by two sets of lattice vectors their rank cannot exceed 4, for which the only compatible non-trivial rotational symmetries are octagonal and dodecagonal. However, as the building blocks of our system are two sets of C6 symmetric moiré lattices, only the dodecagonal quasicrystals can be realized in the absence of strain.

For example in the absence of relaxation, the rigid double moiré structure with $L_{GBN} = L_{GG} = 14.7\ nm$ ($\theta_{GG} = 0.96°$; $\theta_{GBN} = 0°$) where $\vec{K}_{GG}$ and $\vec{K}_{GBN}$ together would form a MQC . But since this point falls within the 1:1 basin of attraction it relaxes into a 1:1 moiré crystal and is thus not observed in the experiment. Outside the 1:1 basin of attraction, two additional MQCs with dodecagonal symmetry can form in the absence of strain. One MQC, observed to the left of the 1:1 line ($L_{GBN} > L_{GG}$ open red circle in Fig. 3) emerges when the $\vec{K}_{GBN}$ vectors together with the moiré-of-moiré vectors $\vec{K}_{MM} = \vec{K}_{GG} - \vec{K}_{GBN}$ form a dodecagon, as shown in the inset of Fig. 2j. The second unstrained dodecagonal MQC, to the right of the 1:1 line ( $L_{GBN} < L_{GG}$ open red star in Fig. 3), emerges when the $\vec{K}_{GG}$ and $\vec{K}_{MM}$ vectors together form a dodecagon. As discussed above, in this region of the phase diagram global alignment is more easily disrupted by external disturbances, which may explain why this MQC was not observed here. Other dodecagonal MQCs can form in the presence of homo or heterostrain of the graphene layers as discussed in Methods.

Table 1 classifies the double-moiré structures with long range order in GG/GBN in terms of periodic moiré crystals, and quasiperiodic crystals (MIC or MQC), according to their rank, translational symmetry, and forbidden rotational symmetry.

| Moiré (M) Type | Rank | | Translational Symmetry? | | Forbidden Rotational Symmetry? | | Equivalent GG/GBN: |
|---|---|---|---|---|---|---|---|
| | 2 | >2 | Yes | No | Yes | No | |
| Crystals | | √ | √ | | | √ | |
| Intercrystals (MIC) | | √ | √ | | | √ | |
| Quasicrystals (MQC) | | √ | √ | | √ | | |

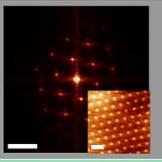
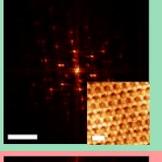

Table 1. **Classification of 2D moiré periodic and quasiperiodic crystals in GG/GBN.** A representative FFT and the corresponding topography (insets) of crystal, MIC and MQC is shown on the right of the table. The right figure schematically illustrates the different moiré structures in GG/GBN. The scale bars are 0.16 nm$^{-1}$ for the FFTs and 16 nm for the topographies.

We next discuss the local electronic properties accessed by STS for the different crystal structures. Fig. 5a, shows the gate voltage ($V_G$) dependence of the dI/dV spectra of a moiré crystal measured at the position of the black star in its topography image (Fig. 2b). $V_G$ controls the sample doping and the Fermi level, which in STM is at bias voltage $V_F = 0$. Similar to magic angle twisted bilayer graphene[41,42], the spectra feature a flat moiré band seen as a prominent peak in the dI/dV spectra. When the Fermi level is outside the flat band, the peak is at its narrowest ~47mV. Upon crossing the Fermi level, the band splits into two peaks separated by a gap signaling an emergent correlated state [41,42]. This is illustrated in more detail in Fig. S12 which shows a maximum gap value of ~40mV. Fig. 5b shows the energy dependent LDoS maps and their FFTs at $V_G = 10V$ for bias voltages from left to right of $V_B = -200$ mV, $-10$mV, and $+200$mV. We find that the electronic wave function preserves the 6-fold symmetry at all measured $V_B$ as illustrated in the FFTs of Fig 5b. We emphasize that the moiré crystals are uniquely described by one set of the well-defined moiré wavevectors $\vec{K}_M$ identified in the zoomed-in FFT of the LDoS map (Fig. 5d) as well as in the FFT of the simulated crystal lattice shown in Fig. 5c. Interestingly, the flat band survives in all the 1:1 moiré crystal studied covering a range of $L_M$ values, 10-15nm, and correspondingly gate voltage spans required to populate the band from empty to full, ranging from 90 V to 60 V(Fig. S13).

Fig. 5e shows the $V_G$ dependent *dI/dV* spectra of an MIC measured at the position of the green star in its topography (Fig. 2h). The spectra reveal a flat band at charge neutrality and a remote flat band at ~+100mV. The former opens a gap (~80mV) as the Fermi level enters it and closes at full filling, ($V_G \sim 10V$), providing a signature of an emergent correlated state as in the crystal case (Fig. S13a). In Fig. 5f, the energy dependent LDoS maps and their FFTs, taken at $V_G = 0V$ and bias voltages $V_B = -10$mV, $+40$mV and $+150$mV, reveal a strong energy dependence of the spatial distribution and symmetry of the electronic wavefunction. We attribute the energy dependent symmetry changes of the LDoS maps to electron scattering off the quasiperiodic GG/GBN potential[43-46]. This is in stark contrast to the energy independent 6-fold symmetry of the moiré crystal maps shown in Fig. 5b.

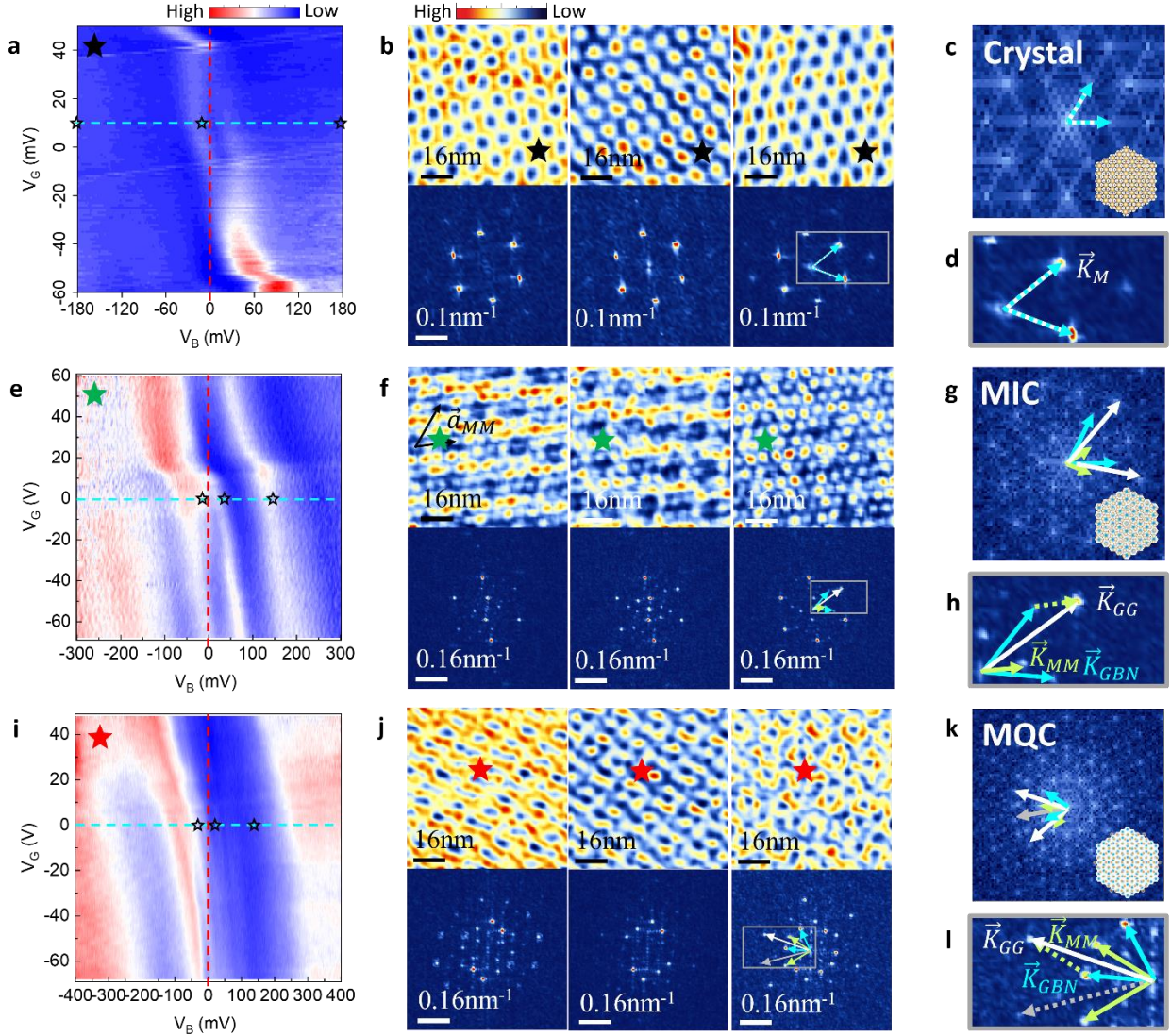

Fig. 5. **Mapping the electronic states of moiré crystal, MIC and MQC. a, e, i,** Gate ($V_G$) dependent STS mapping at locations marked by black, green, and red stars in Fig. 2b, 2h, 2j as well as Fig. 4b, 4f, 4j for crystal, MIC and MQC, respectively, as well as by the stars in panels **b, f** and **j**. The blue dashed lines and open black stars indicate the $V_G$ and bias voltages ($V_B$) of the LDoS maps in panels **b, f** and **j**. **b,** Spatial resolved LDoS mapping of the crystal region in Fig. 2B at $V_B$ = -200 mV, -10 mV, 200 mV are shown in the left, middle and right panels, respectively. **c,** A simulated crystal (lower right inset) and its FFT. **d,** Zoom-in of FFT in **b** lower right panel within the grey rectangular region. **f,** Spatial resolved LDoS mapping of the MIC region in Fig. 2h and the corresponding FFTs measured at $V_B$ = -10 mV, 40 mV, 150 mV **g,** A simulated MIC (lower right inset) and its FFT. **h,** Zoom-in of FFT in **f** lower right panel within the grey rectangular region. **j,** Spatial resolved LDoS mapping of the MQC region in Fig. 2j at $V_B$ = -40 mV, 10 mV, 130 mV are shown in the left, middle and right panels, respectively. Weak peaks at $\vec{K}_{MM}$ and $\vec{K}_{MM} + \vec{K}_{GBN}$ in the lower left panel are amplified by a factor of 3 for clarity. **k,** A simulated dodecagonal MQC (lower right inset) and its FFT. Illustration of the dodecagonal quasicrystal from $\vec{K}_{MM}$ (green arrow) and $\vec{K}_{GBN}$ (blue arrow). **l,** Zoom-in of FFT in **j** lower right panel within the grey rectangular region. The grey arrow represents $\vec{K}_{MM} + \vec{K}_{GBN}$.

Fig 5i shows the doping ($V_G$) dependent *dI/dV* spectra of the MQC sample (solid red star in Fig 3) measured at the position of the red star in its topography (Fig. 2j). The spectra reveal a flat band below charge neutrality which opens a small gap upon approaching the Fermi level. Fig. 5j showcases the STS mappings of the MQC at $V_B$ = -40mV, +10mV and +130mV. Interestingly, the relative intensity of MQC wavevectors is sensitive to energy such that the observed pattern evolves from incommensurate modulated stripes (Fig. 5j left) to checkerboards (Fig. 5j center) and then to dodecagonal (Fig. 5j right). The rigid lattice condition for the dodecagonal MQC constructed by the $\vec{K}_{GBN}$ and $\vec{K}_{MM}$ vectors is given by $\theta_{GG} \approx 2.6°$; $\theta_{GBN} \approx 1°$ and is simulated in Fig 5k. Comparing topography and LDoS maps in Fig. 2j and Fig. 5j (right panel) we conclude that this sample is a dodecagonal MQC which is slightly distorted by strain (Methods). The zoomed in image of Fig. 5j right panel shown in Fig 5l illustrates the formation of the dodecagonal pattern when $|\vec{K}_{GBN}| = |\vec{K}_{MM}| = |\vec{K}_{GG} - \vec{K}_{GBN}|$ and the relative orientation between $\vec{K}_{GBN}$ and $\vec{K}_{MM}$ is 30°.

In summary, we have demonstrated that double moiré potentials, generated by superposing three atomic crystals (graphene-graphene and hBN), give rise to a fascinating array of periodic and quasiperiodic structures with rich emergent electronic properties. These structures revealed a phase diagram featuring lines of commensurate moiré crystals embedded in a sea of quasiperiodic crystals, classified as moiré intercrystals (MICs) or moiré quasicrystals (MQCs) depending on the presence or absence of Bravais-forbidden rotational symmetries.

For rigid lattices, the 1:1 commensurate crystal, which is expected to exhibit orbital magnetism, should theoretically exist at only one point on this phase diagram, rendering it practically undetectable. However, we discovered a self-alignment mechanism that allows commensurate crystals to exist within a basin of attraction extending far beyond the rigid lattice commensuration point. This self-alignment, driven by the competition between the Van der Waals energy gained in favorable AAB stacking and the elastic energy cost of straining the lattice, reduces the sensitivity to twist-angle disorder created during device fabrication.

The double moiré crystals studied here define a new class of quantum materials, providing a synthetic springboard for exploring the electronic properties of quasiperiodic crystals coupled with electron correlations[47], that are expected to host novel forms of magnetism[48], discrete scale invariance[49,50], and exotic superconductivity[51] which are not amenable to analysis by conventional band structure theory. Future progress will benefit from developing theoretical tools for calculating the emergence of flat bands and correlation effects in experimental realizations of quasiperiodic crystals as they are now being observed.

# Methods

## Sample fabrication

Fig S1 shows optical micrographs of the three GG/GBN devices studied (D1, D2 and D3) in this work. Samples were prepared with the tear and stack technique used in our previous work [41]. Thick hBN flakes (30 nm ~ 50 nm) were exfoliated onto a ~300nm $SiO_2$ layer capping an n-doped Si substrate, and monolayer graphene flakes were separately exfoliated onto a membrane of polymethyl-methacrylate (PMMA) 950 A11. The PMMA/graphene stack was then deposited onto a polydimethylsiloxane (PDMS) film with a bump covering a glass slide. The bump on PDMS was made by curing a small droplet of PDMS liquid on a flat PDMS film. The glass slide carrying the graphene-PMMA-PDMS stack was moved to a homebuilt transfer stage within an Argon filled glovebox where the monolayer graphene was deposited on the hBN on $SiO_2$. The graphene and hBN crystals were aligned to the desired twist angle, $\theta_{GBN}$, by collocating their crystal edges. Limited by the resolution of optical microscope and rotation control stage, we estimate the range of $\theta_{GBN}$ variation across the device to be within $-1°$ to $1°$. Following the procedure demonstrated in Fig. S1a, half of the graphene flake was brought into contact with the hBN surface. The PMMA membrane was then lifted at room temperature leaving the contacted half of graphene on the substrate while the other torn-off half stayed with the PMMA. The substrate supporting the graphene on hBN heterostructure was then rotated by $1°$. The torn-off half graphene was then aligned and pressed onto the twisted graphene on hBN. Lifting the PMMA membrane again at room temperature left the second graphene flake which adhered to the first, forming a GG stack with the bottom layer aligned to hBN. For D1, a PMMA shadow mask with electrode window was transferred and aligned over the sample. For D2 and D3, another layer of thin PMMA 950 A6 was spin coated to the surface and with this PMMA layer over the GG/GBN stack we could optically identify the location of GG on hBN. An electrode window was put onto the PMMA around the GG stack through standard e-beam lithography on D2 and D3. All three heterostructures were subsequently contacted by a Ti/Au (4nm/40nm) electrode that was evaporated through the PMMA window using an e-beam evaporation system. The shape of the electrode was customized to allow for capacitive navigation towards the sample region at low temperature[28]. The shadow mask on D1 was removed after metal deposition with scotch tape as a final step before loading it into the STM. D2 and D3 were soaked in acetone overnight for liftoff at room temperature and subsequently annealed in forming gas (10% Hydrogen and 90% Argon) at $300°C$ for more than 24 hours to remove all polymer residues before loading them into the STM. Results in Fig. 2, 5, S2-S7, S12-S13, S15-S21 are from D1 and D2, and Fig. S4j is from D3.

## STM measurement

STM and STS measurements were performed in a homebuilt low temperature high vacuum STM with a base temperature T = 4.5 K using an etched tungsten tip [52] on the D1 and D2 samples. The D3 sample was measured with a mechanically cut PtIr tip in a homebuilt cryogen-free ultra-high vacuum STM at a base temperature of T = 4.6K [53]. The tip was prepared on a gold surface at base temperature and calibrated at the nearby monolayer graphene area on the sample with a V-shape dI/dV spectra for Dirac fermions. The sample was supported on a ~300nm $SiO_2$ layer capping an n-doped Si substrate which serve as a back gate. Gate voltage $V_G$ is applied to the back-gate separated from the sample by the $SiO_2$ and hBN dielectric to tune the sample doping, while bias voltage $V_B$ (-500mV unless otherwise specified) is applied to the sample to maintain a constant tunneling current $I = 20\ pA$ unless otherwise specified. The STM tip was navigated to the sample area using a STM tip-electrode capacitance sensing technique[28]. Regions with different moiré wavelengths ($L_{GG}, L_{GBN}$) are found at various locations by moving the tip across the sample with coarse piezo motors.

**Acknowledgements**

**Funding:**

Department of Energy grant DOE-FG02-99ER45742 (XL, AMC, EYA)

Gordon and Betty Moore Foundation EPiQS initiative grant GBMF9453 (XL, AMC, EYA)

Rutgers University, SAS (GL)

NSF CAREER grant No. DMR-1941569 (JHP)

Sloan Research Fellowship through the Alfred P. Sloan Foundation (JHP)

Aspen Center for Physics where part of this work was performed, which is supported by National Science Foundation grant PHY-1607611 (JHP)

Kavli Institute of Theoretical Physics supported in part by NSF under Grants ~NSF PHY-1748958 and PHY-2309135 (J.H.P. and E.Y.A.).

**Author contributions:**

E. Y. A. conceived and supervised the project.

X. L. fabricated, characterized the samples, and performed STM measurements with input from G. L., A. M. C., E. Y. A..

X. L., G. L., J. P., E. Y. A. analyzed and interpreted the results.

T. T. and K. W. synthesized the hBN crystals.

X. L., G. L., E. Y. A. wrote the paper with input from all authors.

All authors discussed the results.

We thank Paul Steinhardt, Ephthimios Kaxiris and Zhenyuan Zhang for insightful discussions, Eunah Kim and Krishnanand Mallayya for help with machine learning tools, Nikhil

Tilak for help with sample fabrication, Daniele Guerci and Justin Wilson for collaborations on related topics and Shiang Fang for discussions at the early stages of this work.

# Supplementary Information for

**Moiré Periodic and Quasiperiodic Crystals in Heterostructures of Twisted Bilayer Graphene and Hexagonal Boron Nitride**

Xinyuan Lai,[1] Guohong Li,[1] Angela M. Coe,[1] Jedediah H. Pixley,[1,2,3] Kenji Watanabe,[4] Takashi, Taniguchi,[4] Eva Y. Andrei[1]*

∗Corresponding author: eandrei@physics.rutgers.edu

# Contents

**Acronyms**: GG – graphene on graphene; GBN – graphene on hBN; MM – moiré-of- moiré; MB- moiré band; MC – moiré crystal; MIC – moiré intercrystal; MQC – moiré quasicrystal; $V_G$ - gate voltage; $V_B$ bias voltage

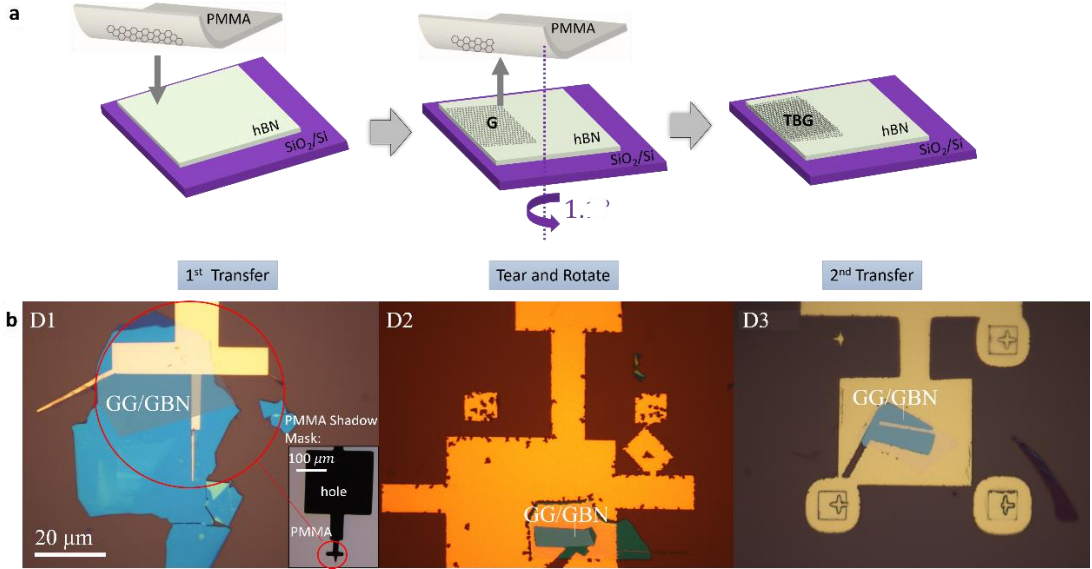

Fig. S1. **Devices for STM measurements. a,** Schematic drawing of the twist heterojunction fabrication procedure. **b,** Device D1, D2 and D3 are presented in the left, middle and right panels, respectively. The inset of the left panel is the optical image of the PMMA shadow mask used to prepare the gold electron. GG/GBN region is overshadowed with grey.

## 1. Identification of stacking order

Fig. S2 shows STM topographies taken at different sample doping and set points of the same region. The triangular GG moiré superlattices characterized by GG AA sites appear higher at full than at empty filling of the flat band in the comparison of Fig. S2a, S2b with S2c, S2d. This is because wavefunction of the GG moiré flat band is strongly localized at the AA sites [1]. The bottom GBN moiré patterns are no longer visible in Fig. S2b but a broken C6 rotational symmetry could be clearly observed from the distortion of GG AB/BA sites. On the contrary, GBN moiré patterns dominate, and we observe mostly hexagonal patterns in Fig. S2c. From the evolution of contrast, we find GG AA sites overlap with GBN $C_B$ sites thus we can confirm the stacking order of this region is AAB. An AAB unit cell consists of a bright spot matching the dark spot is imaged at $V_B$ where both GG and GBN moiré patterns have similar contrasts under our choice of parameters: $V_B$ = -500mV; $V_G$ = 0 V. More examples of 1: 1 commensurate GG/GBN double moiré patterns can be found in Fig. S5.

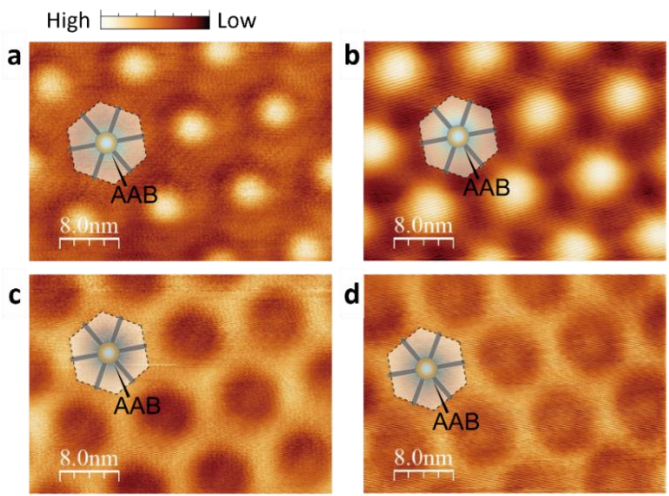

Fig. S2. **Gate and Bias dependence of topographies.** Topography of region with $\theta_{GG} = 1.24°; \theta_{GBN} = 0.79°$ at different $V_B$ and doping: **a,** $V_B = -500mV$; Full Filling ($V_G = 60V$). GG has more contrast than GBN. **b,** $V_B = -200mV$; Full Filling ($V_G = 60V$). GG dominates. **c,** $V_B = -200mV$; Empty Filling ($V_G = -55V$). GBN dominates. **d,** $V_B = -500mV$; Empty Filling ($V_G = -55V$). GBN has more contrast than GG. A schematic drawing of one AAB unit cell is overlapping with one unit cell in each topography. Tunneling current: I = 20pA.

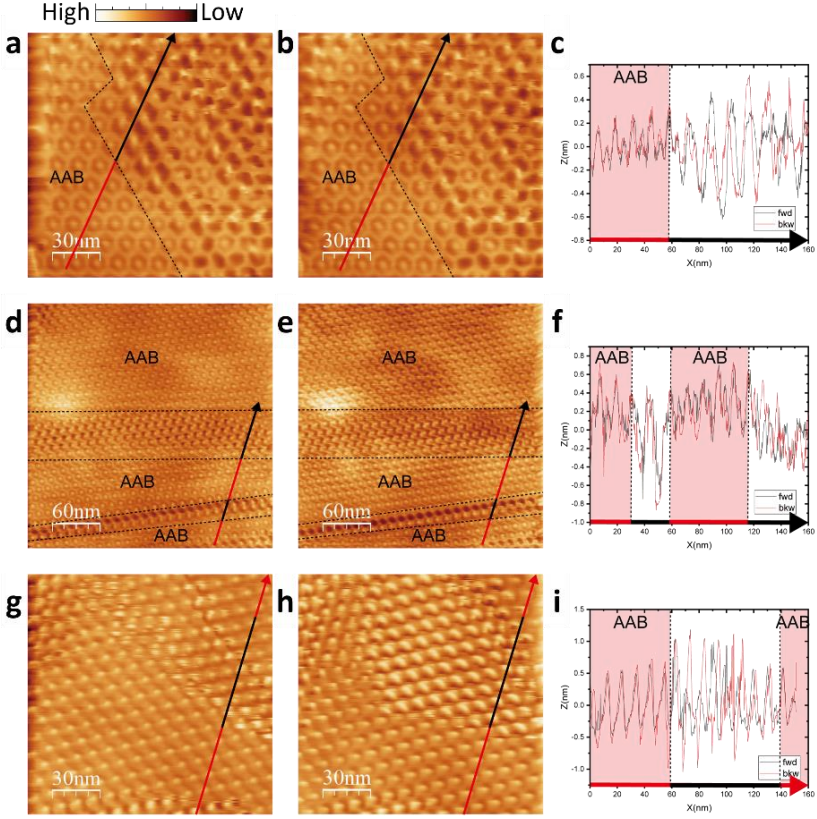

Fig. S3. **Comparing forward and backward STM scans. a,** Forward (left to right) topography scan of region with $L_M = 13.5\ nm$ ($\theta_{GG} = 1.05°; \theta_{GBN} = 0.30°$ for rigid lattices) **b,** Backward (right to left) scan of region in **a**. **c,** Height profile along the arrow in **a** and **b**. **d,** Forward (left to right) topography scan of region with $L_M = 11.0\ nm$ ($\theta_{GG} = 1.28°; \theta_{GBN} = 0.80°$ for rigid lattices) that is large enough to include the two kind of domain boundaries. **e,** Backward (right to left) scan of the same region in **d**. **f,** Height profile along the arrow in **d** and **e**. **g,** Forward (left to right) topography scan of region with $L_M = 10.3\ nm$ ($\theta_{GG} = 1.36°; \theta_{GBN} = 0.93°$ for rigid lattices. **h,** Backward (right to left) scan of the same region in **g**. **i,** Height profile along the arrow in **g** and **h**.

## 2. Stacking order stability

The forward and backward scans presented in Fig. S3 are the same within the AAB stacked domains but are different at the non-AAB regions. This is especially clear in the comparison of height profile linecuts in Fig. S3c, S3f, S3i where AAB stacked regions are identical in the forward and backward profiles while non-AAB stacked regions significantly mismatch in terms of both the phases and magnitudes. This proves that non-AAB stacked regions are less stable under external perturbations from a scanning STM tip. AAB is preferred among the AAB, AAC and AAN stacking orders.

In addition to the 1:1 commensuration, 1:2, 1:3, 1:4, 1:5 commensurations for $\theta_{GG} < 1°$ are also observed in Fig. S4. We note that most GG AA sites align to the nearest GBN $C_B$ sites, forming the preferred local AAB stacking. The spacing between each AAB sites ($L_{GG}$) in these topographies is an integer multiple of $L_{GBN}$. This differs from the smooth wavelength transitions observed in previous works arising from strain and twist angle gradients [2,3], thus further confirms that AAB is the preferred stacking and can form spontaneously. On the other hand, for large GG or GBN twist angles (Fig. S4i: $\theta_{GG} > 1°$; S4j: $\theta_{GBN} > 1°$), there is no self-alignment due to small moiré unit cells are not energetically favorable to relax [4] such that the double moiré patterns remain MIC even when they are very close to commensuration, evidenced by the gradually changing patterns in the left panels of Fig. S4i and S4j.

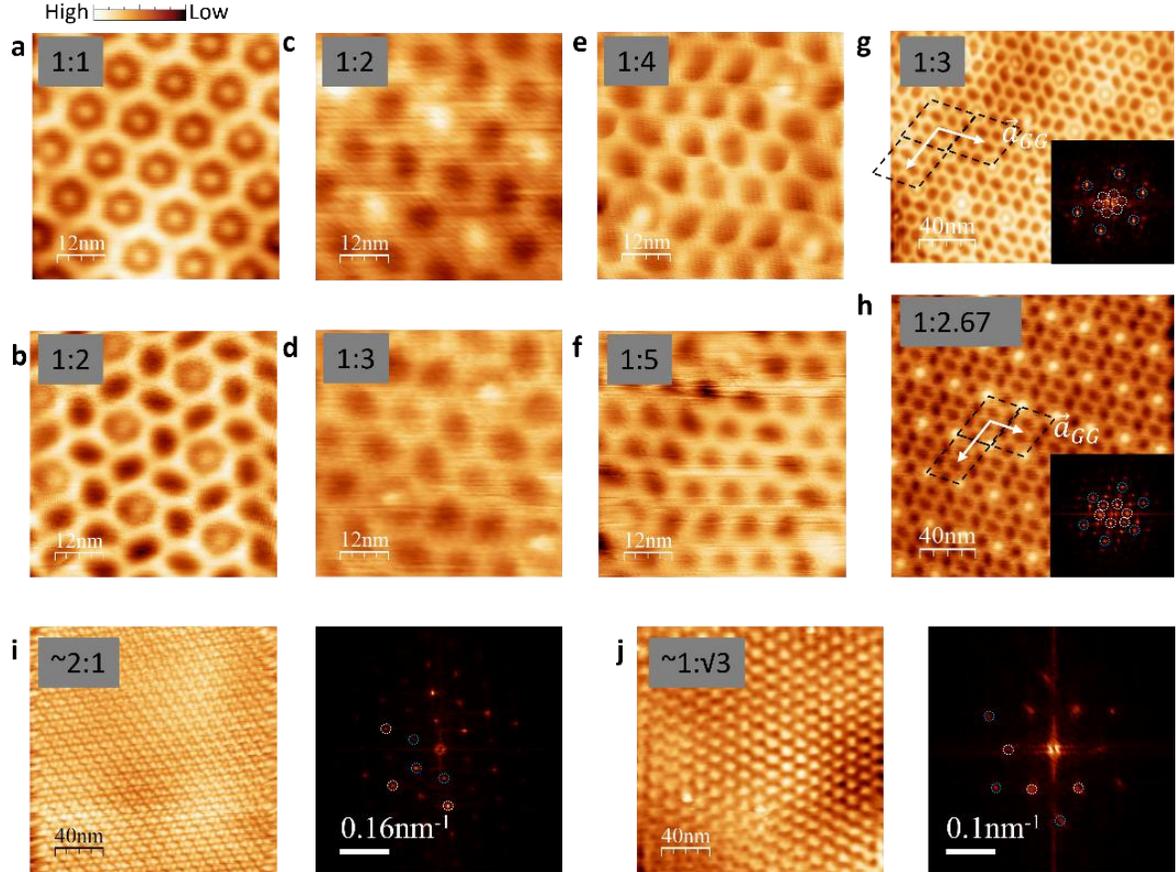

Fig. S4. **Additional commensurate or close to commensurate cases.** The self-alignment leads to many locally commensurate cases at small twist angles for $L_{GG}:L_{GBN} = 1:1, 1:2, 1:2, 1:3, 1:4, 1:5$ in **a,** $L_M = 12.9\ nm$. **b,** $L_{GG} = 21.4\ nm; L_{GBN} = 10.7\ nm$. **c,** $L_{GG} = 27.4\ nm; L_{GBN} = 13.7\ nm$. **d,** $L_{GG} = 38.1\ nm; L_{GBN} = 12.7\ nm$. **e,** $L_{GG} = 44.4\ nm; L_{GBN} = 11.1\ nm$. **f,** $L_{GG} = 50.5\ nm; L_{GBN} = 10.1\ nm$, respectively. **g,** Large commensurate domains are observed for the 1:3 case with $L_{GG} = 35.1\ nm; L_{GBN} = 11.7\ nm$. The inset is FFT of this topography, the window size of FFT is 0.4 by 0.4 (1/nm). **h,** Large commensurate domains where the GG moiré pattern is subject to uniaxial strain to the top graphene layer and homogeneously distorted moiré patterns are observed. Here the average moiré wavelength is $L_{GG} = 32.3.\ nm; L_{GBN} = 12.1\ nm$. The translational symmetry is preserved in this case making it commensurate. The primitive superlattice vectors are marked by white arrows and primitive cells are outlined by dashed diamonds. The insets in **e** and **f** are FFTs of the corresponding topography, the window size of FFT is 0.4 by 0.4 (1/nm). Blue and white circles in FFTs highlight the peak for GBN and GG wavevectors, respectively. and commensuration: **i** This is an example of MIC very close to 2:1 commensuration with $L_{GG} = 6.2\ nm; L_{GBN} = 11.8\ nm$. **j** This is an example of MIC very close to 1: √3 commensuration with $L_{GG} = 12.7\ nm; L_{GBN} = 8.1\ nm$. Note that the $\theta_{GG} = 1.11°$ here is within the magic angle regime. The left panels are topographies, and the right panels are the corresponding FFTs in **i** and **j**. Tunneling parameters: $V_B$ = -1V; $V_G$=0V; I = 30pA for **a** and **b**, $V_B$ = 200 mV; $V_G$= 70V; I = 10pA for **j**, $V_B$ = -500mV; $V_G$=0V; I = 20pA otherwise.

## 3. Topographies of commensurate moiré crystals

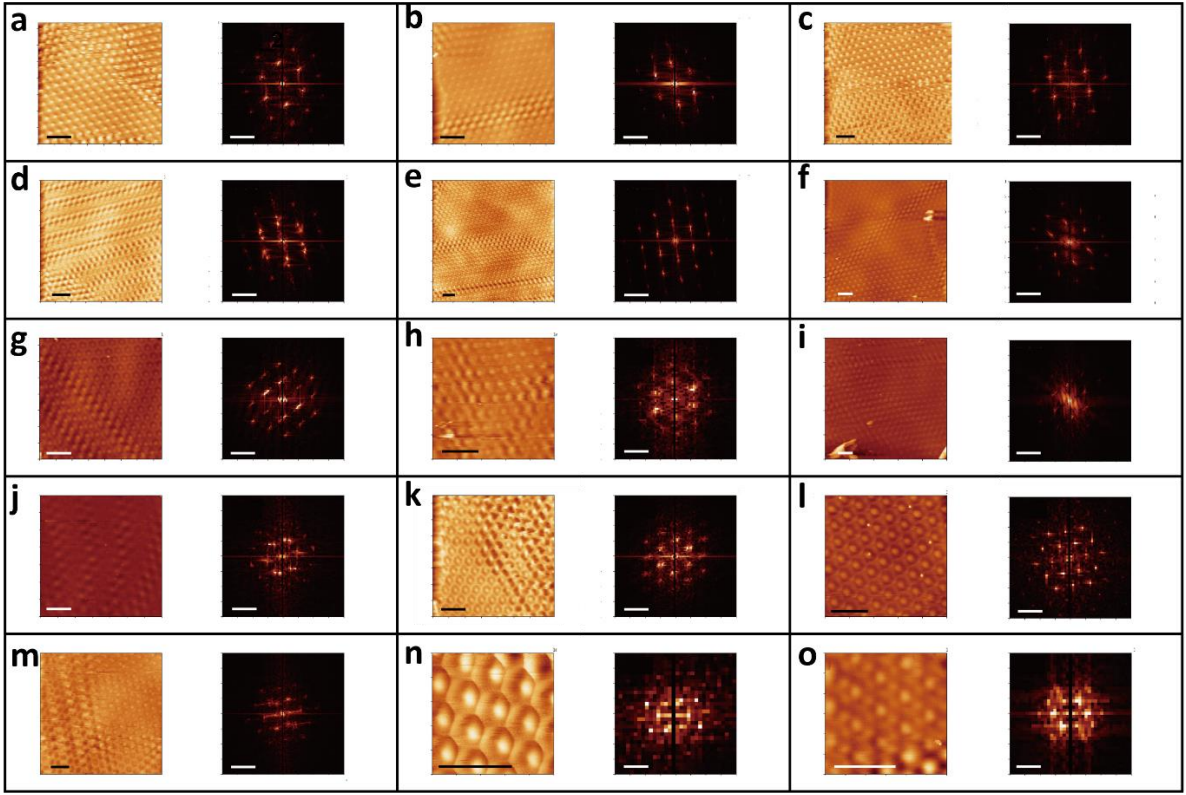

Fig. S5. **STM topographies of $L_{GG}$: $L_{GBN} = 1:1$ commensurate cases with the corresponding FFT.** The peaks at the 2nd Brillion zone have high intensity and are used as indicators for GG/GBN double moiré patterns. The scale bars are 20 nm for topographies (left panels) and 0.8 nm$^{-1}$ (right panels) for FFTs. The topographies are in ascending order of $L_{GG}$.

The examples in Fig. S5 show that AAB stacking is observed to be the only stacking order within the basin of attraction for all 1:1 commensurate regions up to $\theta_{GG} \approx 1.4°$ ($L_{GG} \approx 10\ nm$) in GG-GBN. We find these commensurate AAB regions are either separated by domain boundaries or adjacent to a disordered region with various possible stacking orders. The moiré wavelengths are measured by averaging three crystallographic directions. Fig. S5d is a special case where stripe like domains and sharp "tensile" like boundaries are observed but its wavelength falls outside the basin of attraction. We categorize it as MIC in the phase diagram, but it could be an example of strain assisted self-alignment as an extreme case that worth noting.

The moiré wavelength is measured by the brightest pixel in FFT. The error margins are estimated by the spot size ($1/s$) in FFT and converted back to real space with formula:

$$error = \frac{L}{\left(1 - \frac{\sqrt{3}}{4s}L\right)} - \frac{L}{\left(1 + \frac{\sqrt{3}}{4s}L\right)}.$$

where $L = L_{GG}\ or\ L_{GBN}$ and $s$ is the width of the corresponding topography image. The first term estimates the $L_M$ upper bond of the corresponding FFT spot and the second term is the lower bound. Note that the error from commensurate double moiré patterns ($L_M = L_{GG} = L_{GBN}$) could be half that of incommensurate regions as we rely on the peaks at the second Brillouin zone to calculate the $L_M$ within the commensurate domains. The error bars estimated with this method is added to the phase diagram in Fig. 3 and the error margin is included in Table S1-3 for Fig. S5-7.

| Fig. S5 | $L_{GG}$ (nm) | $L_{GBN}$ (nm) |
|---|---|---|
| a | 10.3±0.3 | 10.3±0.3 |
| b | 10.6±0.3 | 10.6±0.3 |
| c | 10.7±0.2 | 10.7±0.2 |
| d | 10.8±0.3 | 8.8±0.2 |
| e | 11±0.2 | 11±0.2 |
| f | 11.1±0.2 | 11.1±0.2 |
| g | 11.6±0.4 | 11.6±0.4 |
| h | 11.7±0.6 | 11.7±0.6 |
| i | 12.8±0.3 | 12.8±0.3 |
| j | 13.4±0.6 | 12.8±0.5 |
| k | 13.5±0.5 | 13.5±0.5 |
| l | 13.6±0.8 | 13.6±0.8 |
| m | 13.9±0.4 | 14±0.4 |
| n | 14±2 | 14±2 |
| o | 15±2 | 15±2 |

Table. S1 The measured wavelengths from 2$^{nd}$ Brillion zone of Fig. S6 are presented in the Table S1. Since we measure the wavelength of commensurate case at the second Brillouin zone, half the size of the specific pixel at the measured wavelengths translate to real space as $error = \dfrac{L_M}{(2-\frac{\sqrt{3}}{8s}L_M)} - \dfrac{L_M}{\left(2+\frac{\sqrt{3}}{8s}L_M\right)}; L_M = L_{GG}$ or $L_{GBN}$ and $s$ is the width of the corresponding topography image.

## 4. Topography of MICs ($\theta_{GG} > 1°$)

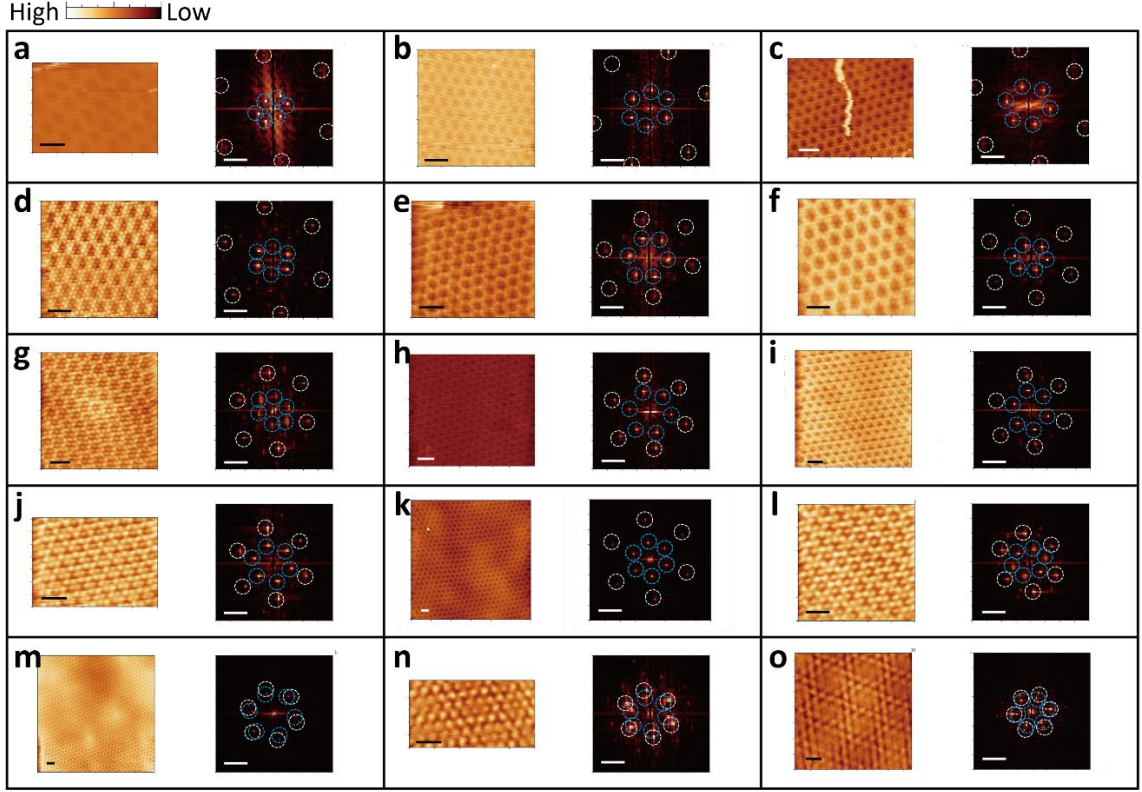

Fig. S6. **STM topographies of MICs and the corresponding FFT ($L_{GG} < L_{GBN}$).** Topographies **(m, n, o)** show very close periodicities in FFT hence the MM is clearly visible. The scale bars are 20 nm for topographies (left panels) and 0.8 nm$^{-1}$ for FFT (right panels). The topographies are in ascending order of $L_{GG}$. Blue and white circles in FFTs highlight the peak for GBN and GG wavevectors, respectively.

We then estimate the possible commensurate length based on the measured wavelength using $L_C \approx L_{GG} * m \approx L_{GBN} * n$. There is always a separation of length scale with $L_C$ being at least one order of magnitude larger than $L_{GG}$ or $L_{GBN}$ for all cases studied here. We thus consider the GG and GBN form quasiperiodic crystals at the moiré scale.

By calculating the ratio $p \equiv L_{GG}\cos\beta/L_{GBN}$ using the two moiré wavelengths, β is the relative angle between the moiré wavevectors. As rational numbers and irrational numbers are practically next to each other in real number set, we could find the closest rational ration of $p$ as $p_r = \frac{a}{b}$; $a$ and $b$ are integers within the range of error, hence find the commensuration length using $L_C = L_{GBN} * b$ and listed them in Table S1. We find this length remains 2 to 3 orders of magnitude larger than $L_{GG}$ or $L_{GBN}$. Such sufficient separation of length scales ($L_C/L_{GBN} \gg 1$) effectively separate the crystals at the length scales of $L_C$ and makes these MICs effective quasiperiodic crystals at the moiré length scales[5].

The expected moiré-of-moiré wavelength ($L_{MM}$) formed between GG and GBN moiré patterns is an indicator of their incommensurability in MICs, namely $L_{MM}$ increases with commensurability and diverge at 1:1 commensuration. The GG reciprocal moiré wavevectors $\vec{K}_{GG0}$ in perpendicular to graphene reciprocal vectors $\vec{K}_{G0}$ under small angle approximation. The angle between $\vec{K}_{GG0}$ and $\vec{K}_{GBN0}$ is thus $\beta \approx atan\frac{\delta}{(\delta+1)\theta_{GBN}} - \frac{\pi}{6}$ where δ is the atomic lattice mismatch between graphene and hBN. With $L_{MM} =$ $\frac{(1+\delta_M)L_{GG0}}{\sqrt{2(1+\delta_M)(1-cos(\beta))+\delta_M^2}}$, we calculate the moiré-of-moiré wavelength for all unstrained GG and GBN wavelength combinations $L_{GG0}$ and $L_{GBN0}$ considering the lattices are rigid. This as well as the conversion relation $L_{GG} \approx \frac{a_G}{\theta_{GG}}$ and $L_{GBN} \approx \frac{a_{BN}}{\sqrt{(\theta_{GBN})^2+\delta^2}}$ gives the plot in Fig. S10a. The $L_{MM}$ characterize the incommensurability near the rigid 1:1 commensurate condition. We then estimate the separation of scales between $L_{MM}$ and moiré length scales ($L_{GG}$ and $L_{GBN}$) in Fig. S10b.

We also find our observation of MIC to consist of twist angles near both 120° and 60° commensuration. Under small angle approximation and assuming the helical rotation is clockwise from the top graphene layer to hBN, the stacking is 120° (helical) if GBN reciprocal lattice vectors are clockwise with respect to GG and is 60° if counterclockwise for 7.4 nm<$L_{GBN}$<12.8 nm (0.55°<$\theta_{GBN}$<1.65°). This rule would be opposite for counterclockwise helical stacking and its anti-helical dual. So even though we clearly see in most examples of Fig. S4 that the GG reciprocal lattice vectors are rotated clockwise compared to GBN. We cannot conclusively determine the helicity because we did not keep track of this during sample preparation and the local rotation is always subject to changes despite the crystal is rotated to a certain angle during fabrication.

| Fig. S6 | $L_{GG}$ (nm) | $L_{GBN}$ (nm) | $L_C$ (nm) |
|---|---|---|---|
| a | 3.33±0.05 | 16±1 | 624 |
| b | 3.59±0.06 | 11.0±0.5 | 2651 |
| c | 3.64±0.05 | 10.2±0.4 | 2591 |
| d | 4.01±0.07 | 13.3±0.8 | 1503 |
| e | 4.9±0.1 | 10.9±0.5 | 1973 |
| f | 5±0.1 | 13.8±0.8 | 3436 |
| g | 5.5±0.1 | 12.8±0.6 | 4262 |
| h | 5.61±0.09 | 9.8±0.3 | 1039 |
| i | 6.0±0.1 | 11.1±0.4 | 5506 |
| j | 6.0±0.2 | 11±0.5 | 4983 |
| k | 7.1±0.1 | 10.9±0.2 | 3150 |
| l | 7.3±0.2 | 14.2±0.9 | 6575 |
| m | 8.04±0.09 | 9.6±0.1 | 3005 |
| n | 8.5±0.3 | 11.2±0.5 | 560 |
| o | 10.7±0.3 | 14.5±0.6 | 8395 |

Table S2. The measured wavelengths from Fig. S7 are shown in Table S2. The size of the pixel at the measured wavelength translate to real space as systematic $error = \frac{L_M}{(1-\frac{\sqrt{3}}{4s}L_M)} - \frac{L_M}{(1+\frac{\sqrt{3}}{4s}L_M)}; L_M = L_{GG}$ or $L_{GBN}$ and $s$ is the width of the corresponding topography image. We find the closest rational number $\frac{n}{m}$ with $n$ and $m$ being integers to the ratio $\frac{L_{GG}*\cos(\beta)}{L_{GBN}}$ for each case where $\beta$ is the relative angle between the GG and GBN, hence calculate $L_C$.

## 5. Topographies of MICs ($\theta_{GG} < 1°$)

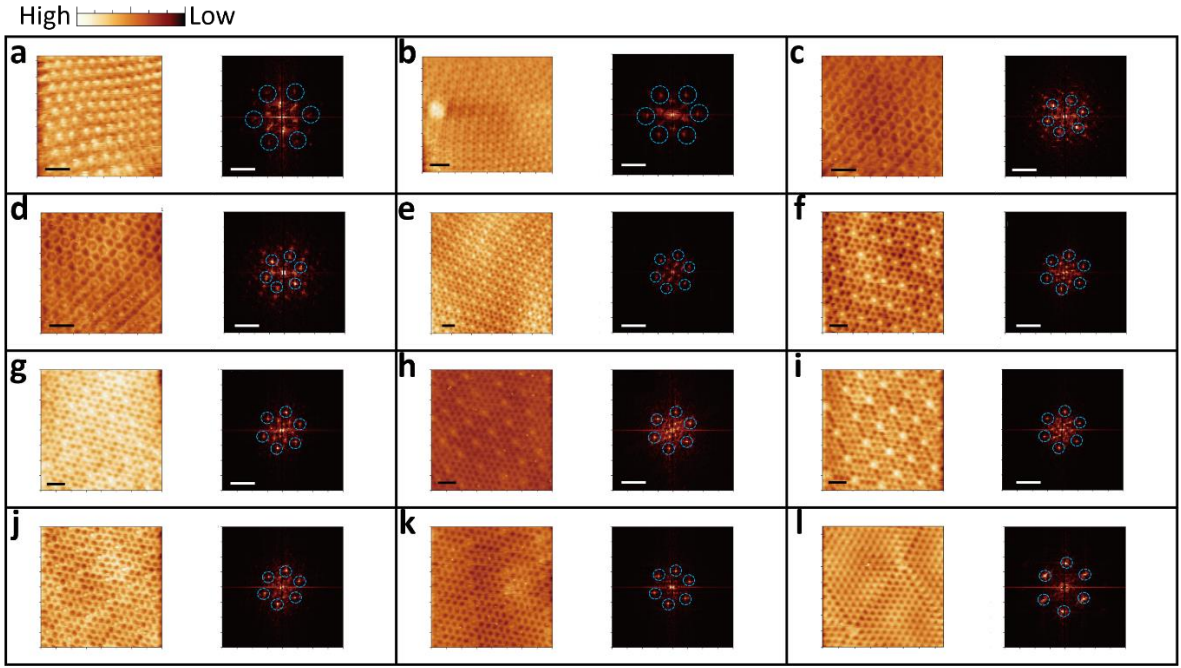

Fig. S7. **STM topographies and the corresponding FFT of MICs ($L_{GG} > L_{GBN}$).** The scale bars are 20 nm for topographies (left panels) and 0.8 nm$^{-1}$ for FFTs (right panels). The topographies are in ascending order of $L_{GG}$. The topographies are in ascending order of $L_{GG}$. Blue circles in FFTs highlight the peak for GBN wavevectors. Due to the strong self-alignment in these data, the FFT peaks for GG wavevectors are not well defined and we typically observe a group of 6 clouds near the center of each FFT for GG.

| Fig. S7 | $L_{GG}$ (nm) | $L_{GBN}$ (nm) |
|---|---|---|
| a | 15.5±0.7 | 7.3±0.2 |
| b | 16.5±0.6 | 9.3±0.2 |
| c | 22±1 | 12.6±0.5 |
| d | 23±1 | 13.1±0.5 |
| e | 27±1 | 12.9±0.2 |
| f | 34±2 | 12.5±0.3 |
| g | 35±3 | 11.2±0.3 |
| h | 36±3 | 11.8±0.3 |
| i | 38±3 | 11.8±0.3 |
| j | 39±3 | 12.7±0.3 |
| k | 61±8 | 12.6±0.3 |
| l | NA | 9.6±0.3 |

Table S3. We measured the moiré wavelengths by identifying the center of the cloud from Fig. S8 and listed them in the Table S3 The size of the specific pixel at the measured wavelengths translate to real space as $error = \frac{L_M}{(1-\frac{\sqrt{3}}{4s}L_M)} - \frac{L_M}{(1+\frac{\sqrt{3}}{4s}L_M)}$; $L_M = L_{GG}$ or $L_{GBN}$ and $s$ is the width of the corresponding topography image. We find the closest rational number $\frac{n}{m}$ with $n$ and $m$ being integers to the ratio $\frac{L_{GG}}{L_{GBN}}$ for each case.

## 6. Derivation of the moiré self-alignment model

To analyze the mechanism behind the observed 1:1 commensurate regime, we assume that only the two graphene monolayers relax, whereas the bulk hBN is rigid. In Fig. S8a-c (or Fig. 4a-c), we show the GG/GBN Brillouin zone corresponding to the 1:1 rigid lattice commensuration condition at $\varphi = 60°$ (or 120°), where $|\theta_{GBN}| = \frac{\theta_{GG}}{2}$. In general, misaligned lattices will have $|\theta_{GBN}| \neq \frac{\theta_{GG}}{2}$ as well as $\varphi \neq 60°$ (or 120°). Global commensuration would thus require rescaling of the lattice constants so that $K_{GG} = K_{GBN} \equiv K_M$, and their relative angle is 60° (or 120°) [6,7]. For the symmetric alignment case, where $\varphi = 60°$ and $|\theta_{GBN}| = \frac{\theta_{GG}}{2}$ (Fig. S8b, S8c top panels) both graphene layers stretch or contract by the same amount corresponding to a reciprocal lattice vector change for both top and bottom graphene layers of $K_G = \frac{\sqrt{3}}{2}(K_{GG0} - K_{GG})$ and to a symmetric atomic strain $\varepsilon_s = \frac{\Delta K_G}{K_{G0}} = \frac{\sqrt{3}}{2}\frac{a_G}{L_M}\left(\frac{\Delta L_M}{L_{M0}}\right)$. For $|\theta_{GBN}| \neq \frac{\theta_{GG}}{2}$ (Fig. S8b bottom panel) the alignment condition is reached through additional amounts of strain added to $\varepsilon_s$ with opposite signs in the two layers, resulting in the asymmetric strain $\varepsilon_a$, as illustrated in Fig. S8b where $\varepsilon_a$ in the top layer is opposite to that of the bottom layer[7]. The elastic energy cost per moiré cell for aligning the two lattices, $E_{el}$, as a function of the twist angles is presented in the next section.

Following the model described by Fig. 4 and Fig. S8a-c, we first solve for the required symmetric lattice strain $\varepsilon_s$ for any combination of $\theta_{GG}$ and $\theta_{GBN}$ as long as $|\theta_{GBN}| = \theta_{GG}/2$ is satisfied. $\varepsilon_s = \frac{K_{G0}}{K_G} - 1 = \frac{a_G}{a_{G0}} - 1$ is the symmetrical lattice strain. $K_G$ is the reciprocal lattice constant of graphene; $K_G'$ is the graphene reciprocal lattice constant after symmetric relaxation; $K_{BN}$ is the reciprocal lattice constant of hBN; $\theta_{GG}$ is the GG twist angle; $\theta_{GBN}$ is the GBN twist angle; $\delta = \frac{K_G}{K_{BN}} - 1 = \frac{a_{BN}}{a_G} - 1$ is the lattice mismatch of graphene and hBN.

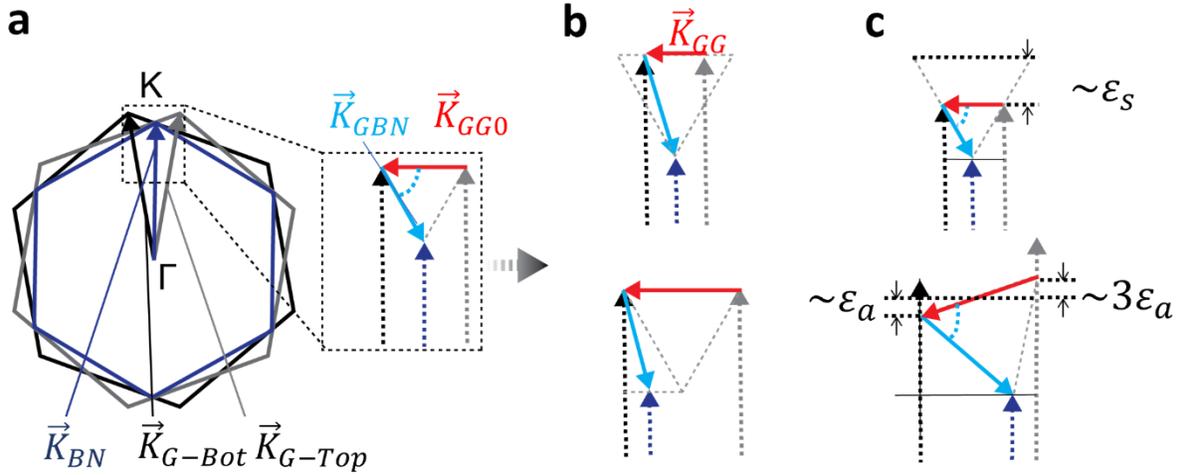

Fig. S8. **Geometric Analysis of Lattice Relaxation for 1:1 moiré crystal near the Magic Angle. a,** Schematic drawings of reciprocal lattice vectors of top and bottom graphene layers ($\vec{K}_{G-Top}$, $\vec{K}_{G-Bot}$) and hBN ($\vec{K}_{BN}$). Zoom-in shows the 1:1 rigid lattice commensurate condition for 60° commensuration. **b,** Schematic drawing of incommensurate cases with $\theta_{GBN} \neq 1.1°$ (top panel) or $|\theta_{GBN}| \neq \frac{\theta_{GG}}{2}$ (bottom panel). **c,** Schematic drawings of self-alignment with symmetrical ($\varepsilon_s$, top panel) or asymmetrical ($\varepsilon_a$, bottom panel) strain.

Since $|\vec{K}_{GG}| = |\vec{K}_{GBN}|$; $\angle(\vec{K}_{GG}, \vec{K}_{GBN}) = 60°$ or $120°$ when GG and GBN moiré patterns are 1:1 commensurate, we can solve the equations: $\frac{2}{\sqrt{3}}(K_G \cos(\frac{\theta_{GG}}{2}) - K_{BN}) = 2K_G * \sin(\frac{\theta_{GG}}{2})$; $K_G = \frac{K_{G0}}{1+\varepsilon_s} = \frac{(1+\delta)K_{BN}}{1+\varepsilon_s}$. This gives: $\varepsilon_s = \left(\cos(\frac{\theta_{GG}}{2}) - \sqrt{3}\sin(\frac{\theta_{GG}}{2})\right)(1+\delta) - 1$. The 1:1 alignment happens near $\theta_{GG}$ and $\theta_{GBN}$ around 1° so we simplify this formula with small angle approximation:

$$\varepsilon_s = \delta - \frac{\sqrt{3}}{2}\theta_{GG}. \tag{4}$$

This formula is identical for both 60° and 120° 1:1 commensuration.

Asymmetric lattice strain $\varepsilon_a$ can be determined from the geometrical relation in the lower panel of Fig. 4b or 3c when applying small angle approximations. We have:

$$\varepsilon_a = \frac{\theta_{GG}}{2\sqrt{3}} - \frac{\theta_{GBN}}{\sqrt{3}}. \tag{5}$$

We also show by comparing Fig. 4b, 4c with Fig. S8b, S8c that 120° and 60° commensuration are geometrically similar where the top and bottom lattice distortions are the combination of symmetric and asymmetric lattice strain terms $\varepsilon_s$ and $\varepsilon_a$. For 120° cases, the asymmetric lattice strain on top layers is the asymmetric lattice strain of bottom layer magnified by three times: $\varepsilon_{bot} = \varepsilon_s + \varepsilon_a$; $\varepsilon_{top} = \varepsilon_s + 3\varepsilon_a$. While for 60° cases, the asymmetric lattice strain is inverted for top and bottom graphene layers such that $\varepsilon_{bot} = \varepsilon_s + \varepsilon_a$; $\varepsilon_{top} = \varepsilon_s - \varepsilon_a$. Since 120° and 60° commensuration cases are indistinguishable under STM topography and they all host AAB, AAN and AAC stacking orders, the topographies studied may include both the cases. We have not observed any region with stable 90° commensuration which would produce two separate periodicities ($L_{GG}: L_{GBN} = \sqrt{3}:2$) in Fourier transformations.

We predict the commensurate moiré wavelength for all twist angle combinations with the relaxed atomic lattices:

$$L_M = \frac{(1 + 2|\varepsilon_a|)(1 + \varepsilon_s - |\varepsilon_a|)a_{G0}}{\sqrt{2(1 + 2|\varepsilon_a|)(1 - \cos(\theta_{GG})) + 4|\varepsilon_a|^2}}. \tag{6}$$

$a_{G0} = 0.246nm$ is the lattice constant of graphene. We plotted the twist angle ($\theta_{GG}$ and $\theta_{GBN}$) dependence of $L_M$ in Fig. S9a for 60° 1:1 commensurate case and in S9b for 120° cases. With $\varepsilon_s$ and $\varepsilon_a$ derived as a function of $\theta_{GG}$ and $\theta_{GBN}$, we simulated the top layer strain ($\varepsilon_{bot}$) and bottom layer strain ($\varepsilon_{top}$) dependence of twist angles in Fig. S9c and S9d for 60° 1:1 commensurate cases and in S9e and S9f for 120° 1:1 commensurate cases.

### 7. Calculation of the elastic energy in self-aligned double moiré patterns

Graphene lattices as well as the GG moiré pattern formed between them preserved the $C_6$ rotational symmetry. We can deduce the elastic potential energy density for strained six-fold symmetric 2D system:

$$\Delta E_{el} = \frac{K}{2}(\varepsilon_{xx} + \varepsilon_{yy})^2 + \mu\left(\left(\frac{(\varepsilon_{xx} - \varepsilon_{yy})^2}{2} + 2\varepsilon_{xy}^2\right)\right). \tag{7}$$

$K$ is the 2D bulk modulus and $\mu$ is the shear modulus. $\varepsilon_{ij}$ is the strain tensor.

For homogeneous stretching/compressing of six-fold symmetric system we have $\varepsilon_{xx} = \varepsilon_{yy} = \varepsilon_{top}$ or $\varepsilon_{bot}$; $\varepsilon_{xy} = 0$, thus the elastic energy density for one strained graphene layer is: $\Delta E_{el} = 2K\varepsilon_{top/bot}^2$. Assuming the strain in the top and bottom graphene layers are independent, the elastic energy per GG moiré unit cell within both top and bottom layers is:

$$E_{el-GG} = \Delta E_{el} * \frac{\sqrt{3}}{2}L_M^2 = \sqrt{3}tYL_M^2(\varepsilon_{top}^2 + \varepsilon_{bot}^2) \tag{8}$$

$Y$ is the Young's modulus and $t$ is the thickness of the 2D material. Considering $Y \sim 0.9\ TPa$ and $t\sim 0.34$ being the thickness one graphene layer [8], we calculate $E_{el-GG}$ as a function of $L_M$ using the formulas given above.

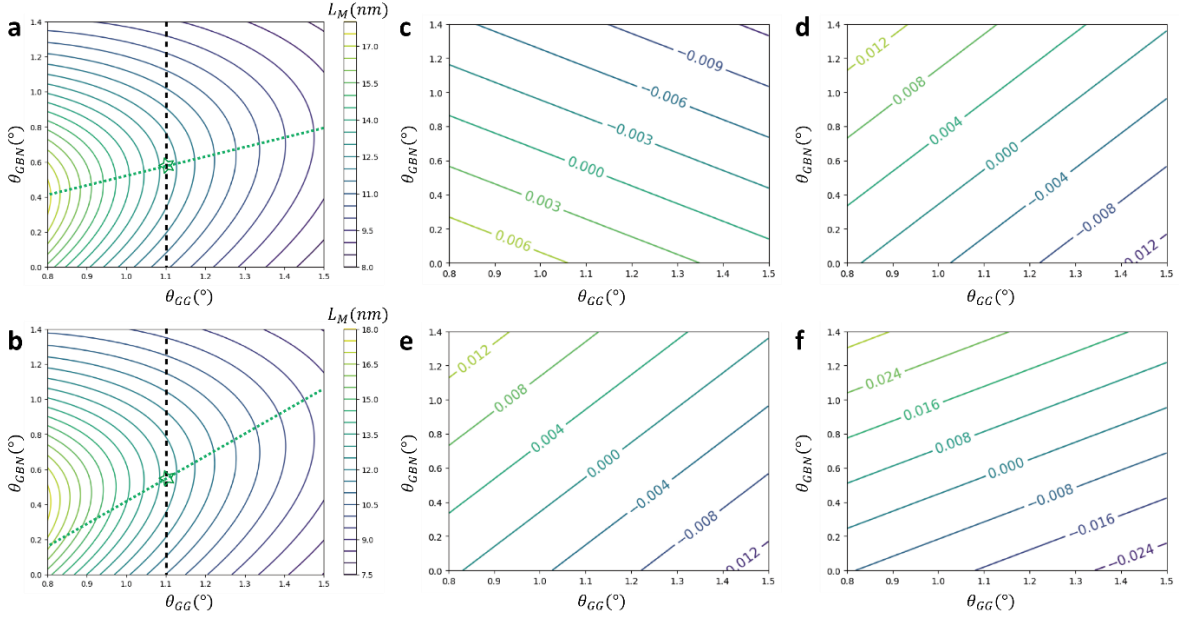

Fig. S9. **Simulated twist angle dependence of moiré wavelength and strain for 1:1 commensurate cases (60° and 120°).** Contour plot of moiré wavelength $L_M$ after self-alignment vs. $\theta_{GG}(°)$ and $\theta_{GBN}(°)$ for the **d,** 60° commensuration and **e,** 120° commensuration. The green line marks the lowest energy cost twist angle combination for a certain $L_M$. The green star marks the rigid lattice commensurate conditions. **f,** Contour plot of bottom layer strain $\varepsilon_{bot}$ vs. $\theta_{GG}(°)$ and $\theta_{GBN}(°)$ for the 60° commensuration. **g,** Contour plot of top layer strain $\varepsilon_{top}$ versus twist angles for the 60° commensuration. **h,** Contour plot of bottom layer strain wavelength $\varepsilon_{bot}$ versus twist angles for the 120° commensuration case. **i,** Contour plot of top layer strain $\varepsilon_{top}$ vs. $\theta_{GG}(°)$ and $\theta_{GBN}(°)$ for the 120° commensuration.

We simulate the dependence of elastic energy per moiré unit cell $E_{el-GG}$ versus $L_M$ and twist angles. This gives the curve in Fig. S11a and 11b for 60° commensurate cases. Elastic energy simulations of 120° cases can be found in Fig. 4d and 4e of the main text. For 60°, the minimal energy cases are simply at $\varepsilon_a = 0$. Under the small angle approximation, we have:

$$E_{el-GG} \approx 2\sqrt{3}tY\left[\frac{\left(1+\delta-\frac{\sqrt{3}}{2}\theta_{GG}\right)a}{\theta_{GG}}\right]^2\left(\delta-\frac{\sqrt{3}}{2}\theta_{GG}\right)^2 \approx 2\sqrt{3}tYL_M^2\left(\delta-\frac{a(1+\delta)}{L_M+\frac{\sqrt{3}}{2}a}\right)^2$$

$$= Yta_G^2\frac{\sqrt{3}}{2}\left[3\left(\frac{L_M}{L_{M0}}-\cos\alpha\right)^2+(\sin\alpha)^2\right]. \tag{9}$$

Which is proportional to $L_M^2$ as expected from the discussion in the main text with small angle approximation and $L_M = \frac{\left(1+\delta-\frac{\sqrt{3}}{2}\theta_{GG}\right)a}{\theta_{GG}}$. We plot the blue curve in Fig. 4f using this equation.

Similarly for the 120° case, we derive:

$$E_{el-GG} = Yta_G^2\frac{\sqrt{3}}{2}\left[3\left(\frac{L_M}{L_{M0}}-\cos\alpha\right)^2+3(\cos\alpha)^2-2\sqrt{3}\left(\frac{L_M}{L_{M0}}-\cos\alpha\right)\cos\alpha+1\right]. \tag{10}$$

which leads to the green curve in Fig. 4f.

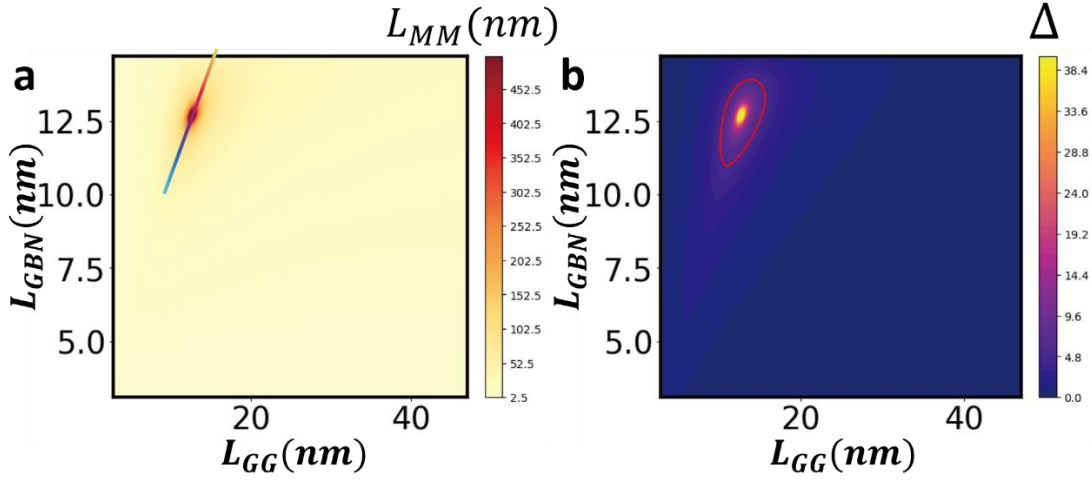

Fig. S10. **Simulated moiré wavelength dependence of moiré-of-moiré wavelength and separation of scale. a,** Simulated moiré-of-moiré wavelength $L_{MM}$ as a function of [$L_{GG0}$, $L_{GBN0}$] assuming graphene and hBN lattices are rigid. The $L_{MM}$ is maximized at the theoretical 1:1 commensurate twist angle combination. The colored line in **j** indicate the range of $L_{MM}$ simulated in S11a, S11b. **b,** Estimated separation of scale $\Delta = \frac{L_{MM}}{Max[L_{GG0},\ L_{GBN0}]}$ versus $L_{GG0}$ and $L_{GBN0}$. We used small angle approximation in calculating $L_{GG0}$ and $L_{GBN0}$ thus the same plot also applies to 120° commensuration. The red circle is $\Delta = 5$.

The preferred stacking of GG-GBN is AA aligned to $C_B$ (AAB). This lowers the local van der Waals (vdW) energy per atom by up to $0.018\ eV$ compared to non-aligned AA sites. The estimated vdW energy gain of AAB aligned GG AA sites is $U \approx p * 0.010 eV$ [9] per unit cell, where $p$ is the number of atoms per AA site. The experimentally measured diameter of GG AA sites ($d_{AA}$) [4] is from 4.6 to 5.4 nm for $\theta_{GG}$ from 1.4° to 0.94° or experimentally observed $L_M$ ranging from 10 to 15 nm. We calculate the number of atoms in an AA site using:

$$p = \frac{\pi d_{AA}^2}{\sqrt{3} * a^2}. \tag{11}$$

This gives the range of $U$ to be from $6.4eV$ to $8.8eV$ for $L_M = 10\ nm$ to 15 nm. Connecting these two points gives the pink line in Fig. 4f as an upper bound estimate of the self-alignment energy.

The elastic energy cost per aligned moiré region is balanced by the Van der Waals energy gain of AAB stacking, U, so that self-alignment will occur for U > E$_{el}$. We obtain a rough estimate of U based on the area of the moiré unit cells. An estimate of the experimental upper bound of stability can be expressed in terms of $E_B$, the vdW energy gain per unit cell in the AAB configuration compared to surrounding non-aligned sites. We obtain $E_B$ ~53 meV per atomic unit cell for the experimentally observed commensurate cases at 60° and $E_B$ ~10 meV for the 120° configuration.

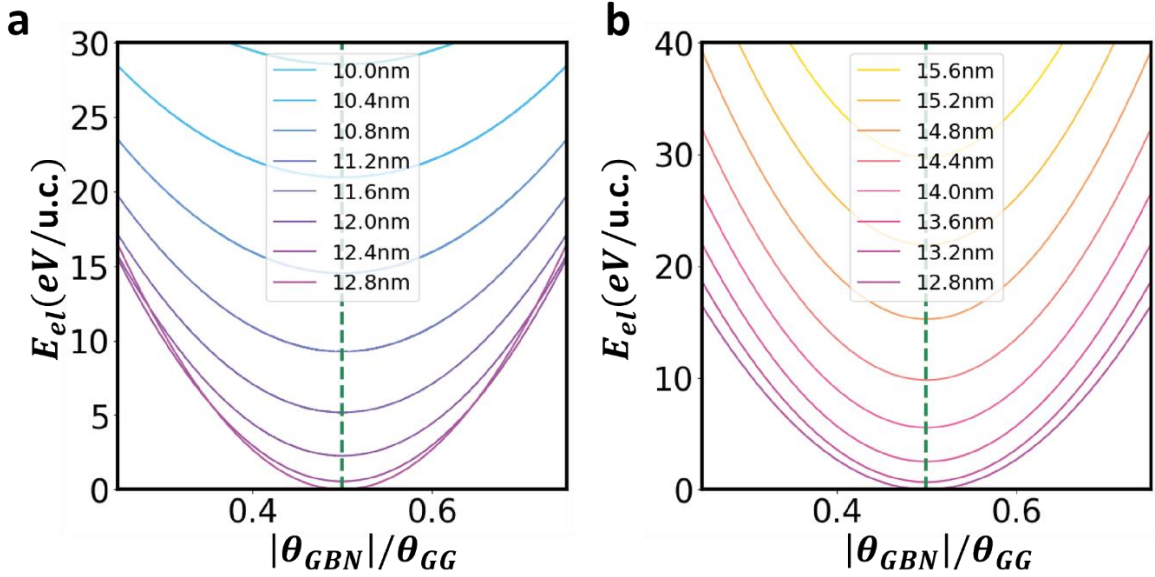

Fig. S11. **Simulated moiré wavelength dependence of elastic energy for 60° 1:1 commensuration.** Elastic energy ($E_{el}$) at a constant wavelength ($L_M$) for different twist angle combinations ($|\theta_{GBN0}|/\theta_{GG}0$) for 60° 1:1 commensurate cases is plotted in **a,** for $L_M < 12.8\ nm$. **b,** for $L_M > 12.8\ nm$.

## 8. Correlated States in GG/GBN

We observe a single band develop into two Hubbard bands with decreasing $V_G$ in the moiré crystal, MIC and MQC, evidenced by a peak split to two peaks near the Fermi level in Fig. 5a, 5e, 5i and Fig S12, and we measure the bandwidth. Correlated electrons localized to disorders, magnetic field or moiré periodical potential are proved to induce soft Coulomb gaps in tunnelling spectroscopies [10-12]. We estimate this Coulomb gap ($\Delta E_c$) by the splitting of flat bands (FB) in Fig. S12d [13], consistent with previous studies [12,14]. This gap does not fully open due to $U$ is comparable to the bandwidth in this non-magic angle moiré crystal region. We do observe fully developed fractional filling states in the FBs at magic angle moiré crystal regions in Fig. S13a-d. The LDoS peak representing the group of MBs in MIC also split unexpectedly near the Fermi level with spectrum weight transfer between the upper/lower Hubbard band peaks from full to partial filling, implying electron correlation driven by the Coulomb interaction. We estimate $U \approx 20\ meV$ from the broadening of the MB peak in this region (Fig. S12e). However, the splitting near the Fermi level indicates a soft gap about $80\ meV$, marking a crucial difference between quasiperiodic MIC and moiré crystal. The inconsistency of the broadening of MB and band splitting is inviting future theoretical studies about the detailed electronic structures of these quasiperiodic crystals. Broadening of MB peak is also observed in the MQC region (Fig. S12f). The sub-band splitting near the Fermi level is measured to be about $30\ meV$, comparable to $U$.

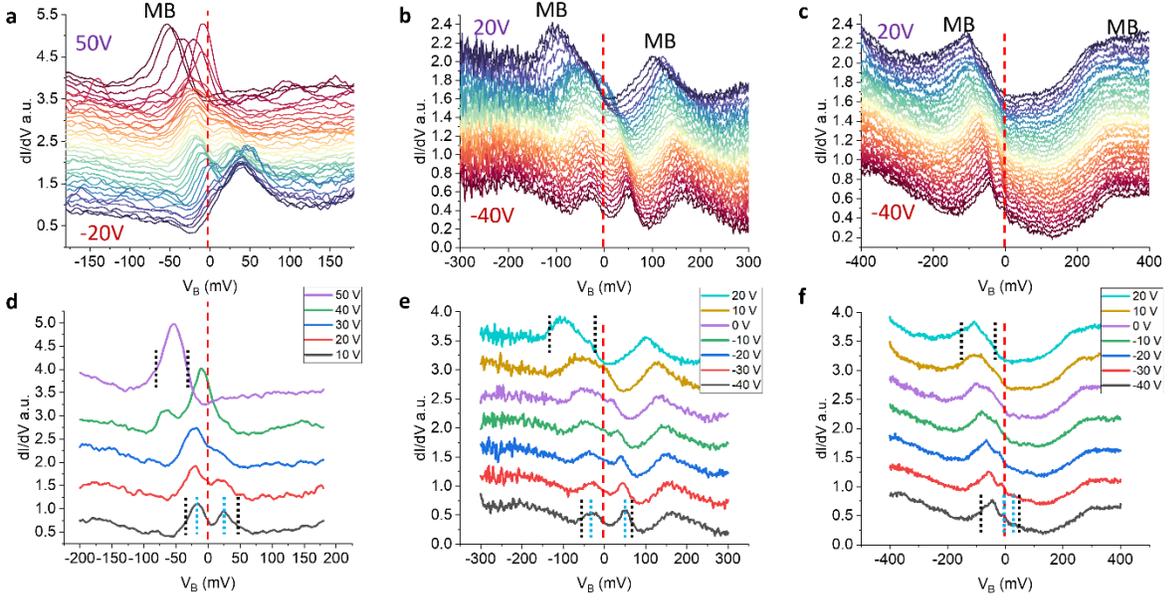

Fig. S12. **Coulomb interaction of** moiré crystal**, MIC and MQC a-c,** Cascade plots of gate ($V_G$) dependent dI/dV at $V_G$ range indicated by the color-coded labels for moiré crystal, MIC and MQC in Fig. 5a, 5e and 5i, respectively. **d-f,** Plots of representative dI/dV spectra from full to partial filling of the MBs extracted from **a-c** at $V_G$ indicated in the legends. We may estimate the Coulomb interaction strength $U$ by the broadening of MB, whose band width ($W$) can be measured from the difference of $V_B$ interval between the black dashed lines, from full filling (high $V_G$) to partial filling (low $V_G$). Alternatively, one may estimate the (soft) Coulomb gap $\Delta E_c$ by the interval between blue dashed vertical lines at empty filling (low $V_G$). We have moiré crystal: $W = 47 \pm 5\ mV$; $\Delta E_c = 40 \pm 10\ meV$; $U = 35 \pm 10\ meV$, MIC: $W = 100 \pm 5\ mV$; $\Delta E_c = 80 \pm 10\ meV$; $U = 20 \pm 10\ meV$, MQC: $W = 120 \pm 5\ mV$; $\Delta E_c = 30 \pm 10\ meV$; $U = 30 \pm 10\ meV$. $\Delta E_c$ is close to $U$ in moiré crystal and MQC while these values deviate substantially in MIC. Tunneling parameters: $V_B$ = -200mV; I = 100pA and $V_B$ = -300mV; I = 100pA for **c** and **f**.

The gate ($V_G$) dependent dI/dV spectra at different moiré crystal regions are presented in Fig. S13a-d. We observe the signature of correlated states at integer electron fillings, reflected by the cascades and transitions of electronic states with doping in Fig. S13b-d for $L_M = 15\ nm, 13.6\ nm$ and $12.8\ nm$, which translates to twist angles $\theta_{GG} = 0.94°, 1.04°$ and $1.1°$ for rigid lattices. This matches the range of the magic angle GG without alignment to hBN [15]. The MB remains relatively flat for the moiré crystal s as evidenced by the dI/dV spectra at full filling without electron correlation in Fig. S13a.

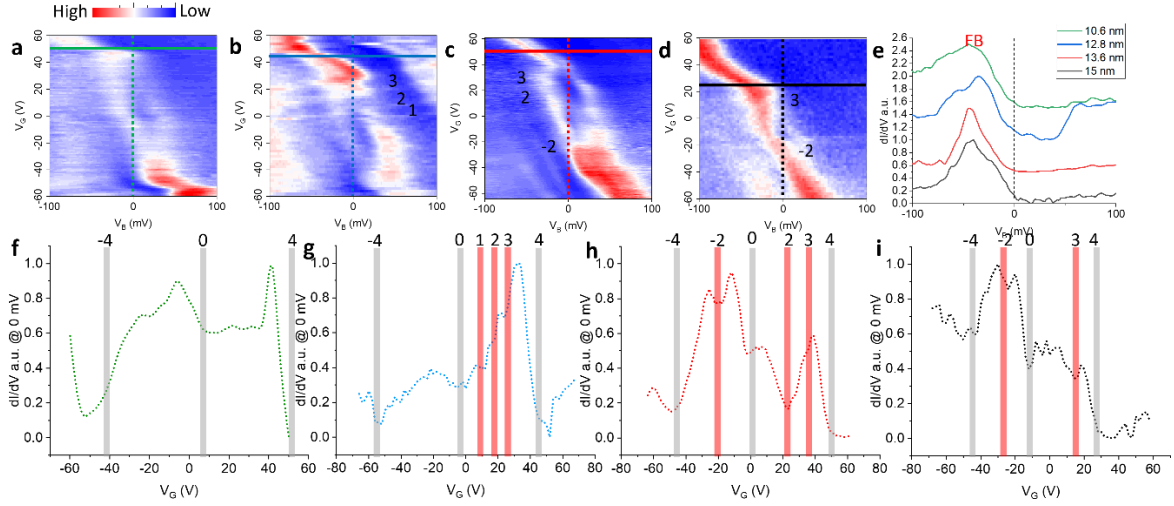

Fig. S13. **Characterization of wavelength dependent electron correlation of moiré crystal. a-d,** Gate dependent STS mapping at AAB site of moiré crystal regions with $L_M = 10.6\ nm$, (Fig 2b), 12.8 nm (Fig. 2a), 13.6 nm (Fig. S5l), 15 nm (Fig. S5o) respectively. Signature of correlated states could be observed in **b-d** with the filling number for each cascades marked, but is missing in **a**. **e,** dI/dV at full filling of the FB extracted from **a-d**. The $V_G$ of the dI/dV curves is marked by the colored lines in **a-d**. **f-i,** Linecuts of **a-d** along the Fermi level/$V_B = 0\ mV$ marked by the vertical dashed lines. The plots show clear dips (vertical shaded bars are guides to the eye and fillings are marked) at integer fillings. Here "-" correspond to the hole doped sectors. dI/dV spectra in **f-i** are normalized to (0, 1). Tunneling parameters: $V_B$ = -200mV; I = 100pA; $V_B$ = -100mV for **a-c**; I = 100pA for **d.**

## 9. Rigid lattice MQC conditions

As discussed in the main text, the MQC formed by the lattice vectors of GG and GBN at $L_{GG} = L_{GBN} = 14.7\ nm$ falls within the self-alignment basin of attraction, and hence relaxes to a 1:1 moiré crystal. The second order MQCs from GBN (or GG) and MM, the difference of GG and GBN, fall outside the basin of attraction and may exist in devices. The rigid lattice condition for second order dodecagonal MQCs implies that adjacent GBN (or GG) lattice vectors are rotated by 30 degrees with respect to the MM lattice vector. Recalling that the angle between $\vec{K}_{GG0}$ and $\vec{K}_{GBN0}$ is $\beta \approx atan\frac{\delta}{(\delta+1)\theta_{GBN}} - \frac{\pi}{6}$, where $\delta$ is the atomic lattice mismatch between graphene and hBN, and solving for $\beta = 15°$ we get $\theta_{GBN} = 0.98°$ and $L_{GBN} = 10.3\ nm$. The second condition requires $|\vec{K}_{MM0}| = |\vec{K}_{GG0} - \vec{K}_{GBN0}|$ equals to $|\vec{K}_{GG0}|$ or $|\vec{K}_{GBN0}|$. The geometric relations in Fig. S14a translates to $L_{GG} = \frac{L_{GBN}}{2\cos(\beta)} = 5.3\ nm$ for MM-GBN MQC, or $L_{GG} = \frac{(2+\sqrt{3})L_{GBN}}{2\cos(\beta)} = 19.9\ nm$ in Fig. S14b for MM-GG MQC. These gives the MQC conditions marked in Fig. 3 by red empty stars. Homo or hetero- straining the lattices deviate from these angle combinations leads to (partial) MQCs as long as the wavevectors satisfy the geometric relation defined in Fig. S14. The example of a slightly strained MQC formed by the wavevectors of GBN and MM is observed experimentally, Fig. 2i, at the red solid star in Fig. 3. We explain its origin in the next section. The observation in Fig. S4j suggests there is no self-alignment near this twist angle combinations, suggesting the existence of MM-GG MQC.

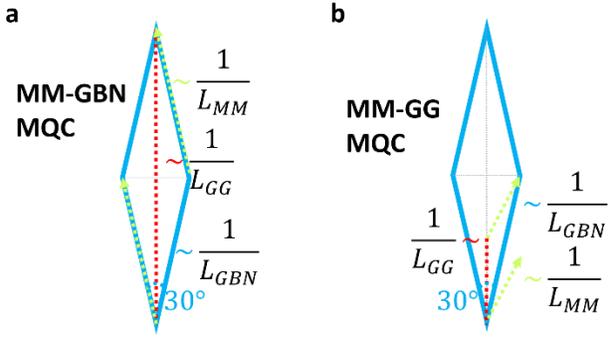

Fig. S14. **Schematic drawings of rigid lattice MQC conditions. a,** MM-GBN or **b,** MM-GG. The inner angle of the diamond is 30 and 150 degrees.

## 10. Analyzing the heterostrain in double moiré patterns

The large area topographies for GG and GBN unit cell examples shown in Fig. 1e is presented in Fig. S15a and S15b, respectively. We identify GG and GBN periods from their continuity (translational symmetry) and difference in the patterns. For MICs and MQC, GG and GBN periods are well separated in FFT. We may analysis the uniaxial strain of the graphene layers in MIC or MQC by measuring the moiré wavelengths in the three of crystallographic directions by a heterostrain model.

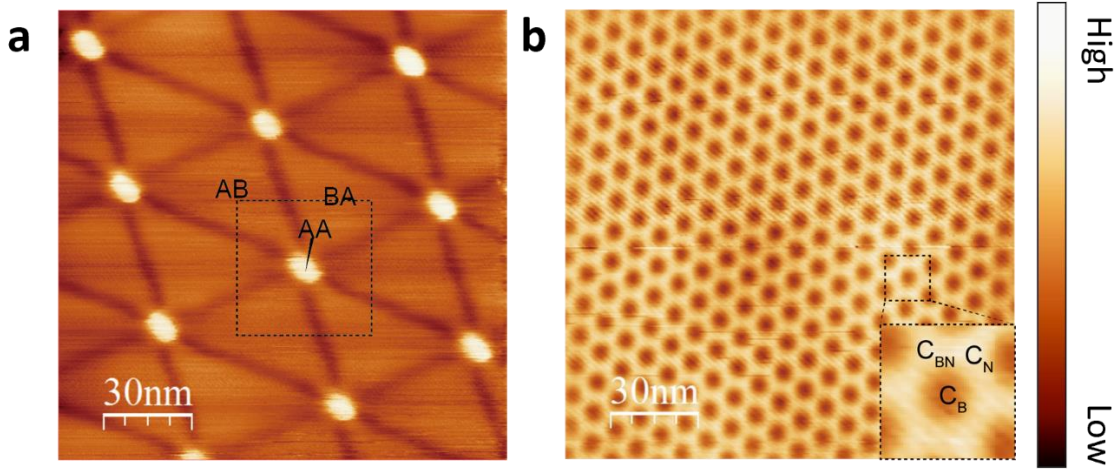

Fig. S15. **Topography of GG and GBN moiré patterns without alignment. a,** Topography of a small angle GG with $\theta_{GG} = 0.28°$ not aligned to hBN. Different regions are marked as AA, AB and BA. **b,** Topography of Monolayer Graphene-hBN moiré patterns as hexagons with $\theta_{GBN} = 1.42°$. Different regions are marked as $C_B$, $C_{BN}$ and $C_N$ in the inset. Topography scans are acquired with $V_G = 0$; $V_B = -300mV$; $I = 20pA$

Considering the coexistence of tensile strain $\epsilon_t$, shear strain $\epsilon_s$, the angle of strain applied $\theta_s$, twist angle of two twisted layers $\theta_t$, we construct the model with the rotated 2D strain tensor operate on the bottom layer graphene reciprocal lattice vectors $\vec{k}_{G-Bot;n}$; $n = 1, 2, 3$ represent the three crystallographic direcitons as schematically illustrate in Fig. S16a similar to ref [16]:

$$\vec{k}'_{G-Bot;n} = \boldsymbol{R}^{-1}(\theta_s) \cdot \boldsymbol{S} \cdot \boldsymbol{R}(\theta_s) \cdot \vec{k}_{G-Bot;n} =$$
$$\begin{pmatrix} cos(\theta_s) & sin(\theta_s) \\ -sin(\theta_s) & cos(\theta_s) \end{pmatrix} \begin{pmatrix} \epsilon_{t;x} & \epsilon_s \\ \epsilon_s & \epsilon_{t;y} \end{pmatrix} \begin{pmatrix} cos(\theta_s) & -sin(\theta_s) \\ sin(\theta_s) & cos(\theta_s) \end{pmatrix} \vec{k}_{G-Bot;n} \qquad (12)$$

We may diagonalize strain tensor such that. $\mathbf{S} = \mathbf{T}^{-1}\mathbf{S'T}$. Thus, we get:

$$\vec{k}'_{G-Bot;n} = \mathbf{R}^{-1}(\theta_s) \cdot \mathbf{T}^{-1} \cdot \mathbf{S'} \cdot \mathbf{T} \cdot \mathbf{R}(\theta_s) \cdot \vec{k}_{G-Bot;n} = \mathbf{R}^{-1}(\theta_s') \cdot \mathbf{S'} \cdot \mathbf{R}(\theta_s') \quad (13)$$

Finally:

$$\vec{k}'_{G-Bot;n} = \begin{pmatrix} \cos(\theta_s') & \sin(\theta_s') \\ -\sin(\theta_s') & \cos(\theta_s') \end{pmatrix} \begin{pmatrix} \frac{1}{1+\epsilon} & 0 \\ 0 & \frac{1}{1-\delta\epsilon} \end{pmatrix} \begin{pmatrix} \cos(\theta_s') & -\sin(\theta_s') \\ \sin(\theta_s') & \cos(\theta_s') \end{pmatrix} \vec{k}_{G1n} \quad (14)$$

We have $\delta$ to be the Poisson ratio graphene.

We may rotate the hBN relative to bottom graphene layer and get moiré reciprocal vector:
$\vec{K}_{GBN;n} = \vec{k}'_{G-Bot;n} - \mathbf{R}(\theta_t) \cdot \vec{k}_{BN;n}$.

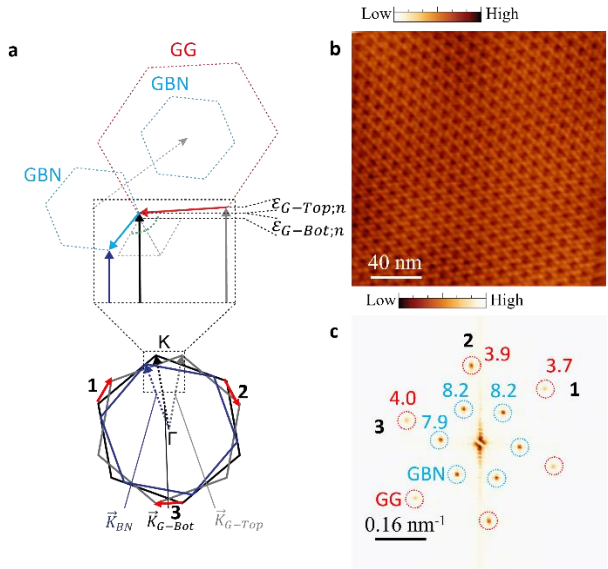

Fig. S16. **Amplified FFT of MQC with $L_{GG}$ and $L_{GBN}$ marked in each of the crystallographic directions. a,** Schematic drawing illustrates the atomic strain ($\varepsilon_{G-Top;n}$, $\varepsilon_{G-Bot;n}$) effect over the moiré lattices in crystallographic direction $n = 3$. **b,** Large topography of the MQC area shown in Fig. 2j. **c,** FFT of **b** with moiré wavelengths marked for either GG or GBN in three of the crystallographic directions. The topography scan is acquired with $V_G = 0$; $V_B = -500\text{mV}$; $I = 20\text{pA}$.

We consider hBN substrate to be rigid as discussed above and the lattice constant of graphene to be 2.46Å, lattice constant of hBN to be 2.493Å. We may calculate the uniaxial strain in bottom graphene layer ($\varepsilon_{bot}$) then use this as a parameter to get the strain in the top graphene layer ($\varepsilon_{top}$) as Fig. S16a shows. From the topography and FFT of the MQC region, we measure moiré wavelengths in the three crystallographic directions following the reciprocal lattice vectors in Fig. S16c. We numerically fit these to equation (14) and get $\varepsilon_{G-bot} = 0.3\%$; $\theta_{S-Bot} = 9.0°$; $\theta_{GBN} = 1.29°$ for bottom layer graphene.

With the above parameters for bottom graphene layer, we apply the uniaxial strain analysis to the bottom and top graphene layers and numerical fit:

$$\vec{K}_{GG;n} = \mathbf{R}^{-1}(\theta_s') \cdot \mathbf{S'} \cdot \mathbf{R}(\theta_s') \cdot \vec{k}_{G-Bot;n} - \mathbf{R}(\theta_t) \cdot \vec{k}'_{G-Top;n} \quad (15)$$

yield $\varepsilon_{G-Top} = 0.5\%$; $\theta_{S-Top} = -19.6°$; $\theta_{GG} = 3.16°$ for top layer graphene. This strain configuration ($\varepsilon_{G-Bot} = 0.3\%$; $\theta_{S-Bot} = 9.0°$; $\varepsilon_{G-Top} = 0.5\%$; $\theta_{S-Top} = -19.6°$) magnifies $\vec{K}_{GG;3}$ while rotates and reduces $\vec{K}_{GBN;3}$, matching them to the MQC condition.

## 11. Additional LDoS Mappings

We present another MIC with crystallographic directions of GG, GBN and the consequent MM aligned in Fig S17. The condition for such alignments is $|\theta_{GBN}| = 0.55°$. The moiré-of-moiré-of-moiré (MMM) pattern as a second order interference of GBN and MM is observed for a wide range of energies, reflecting the incommensurability and scaling symmetry of such quasiperiodic structures. Additional STS mapping data at different energies from the moiré crystal, MIC and MQC regions in main text Fig. 5 and Fig. S17 is presented in Fig. S18-S21.

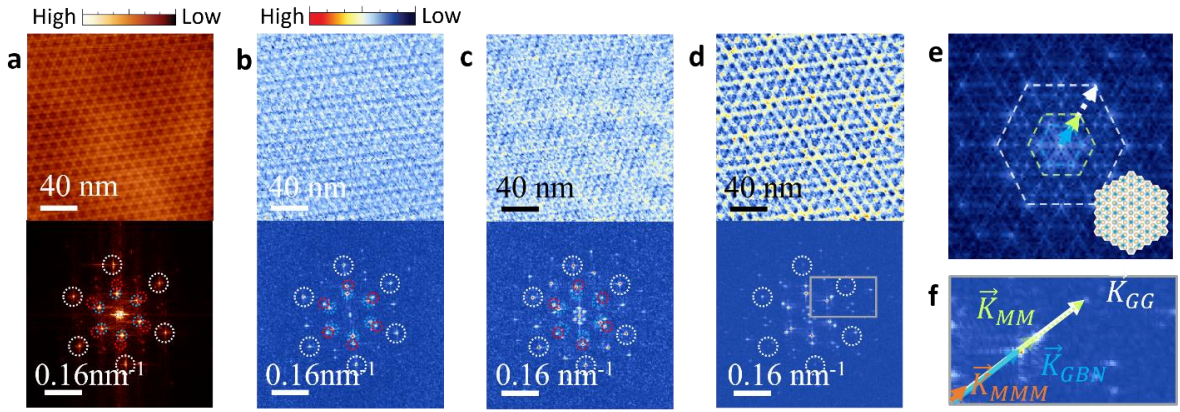

Fig. S17. **LDoS mapping of a MIC with the directions of $\vec{K}_{GG}$ and $\vec{K}_{GBN}$ aligned. a,** Topography of an example of incommensurate modulated crystal: $L_{GG} = 5.43\ nm$; $L_{GBN} = 12.47\ nm$ ($\theta_{GG} = 2.58°$; $\theta_{GBN} = 0.56°$). LDoS mapping of this region at energies $V_B$ = -120mV, -80mV, -40mV are shown in **b-d**, respectively. **e,** A simulated aligned MIC (lower right inset) and its FFT. **f,** Zoom-in of FFT in **d**, within the grey rectangular region. The GG, GBN, MM and 2$^{nd}$ order interference induced moiré-of-moiré-of-moiré (MMM) wavevectors are illustrated with white, blue, green and yellow arrows, respectively. Tunneling parameters: $V_B$ = -200mV; $V_G$=0V; I = 100pA

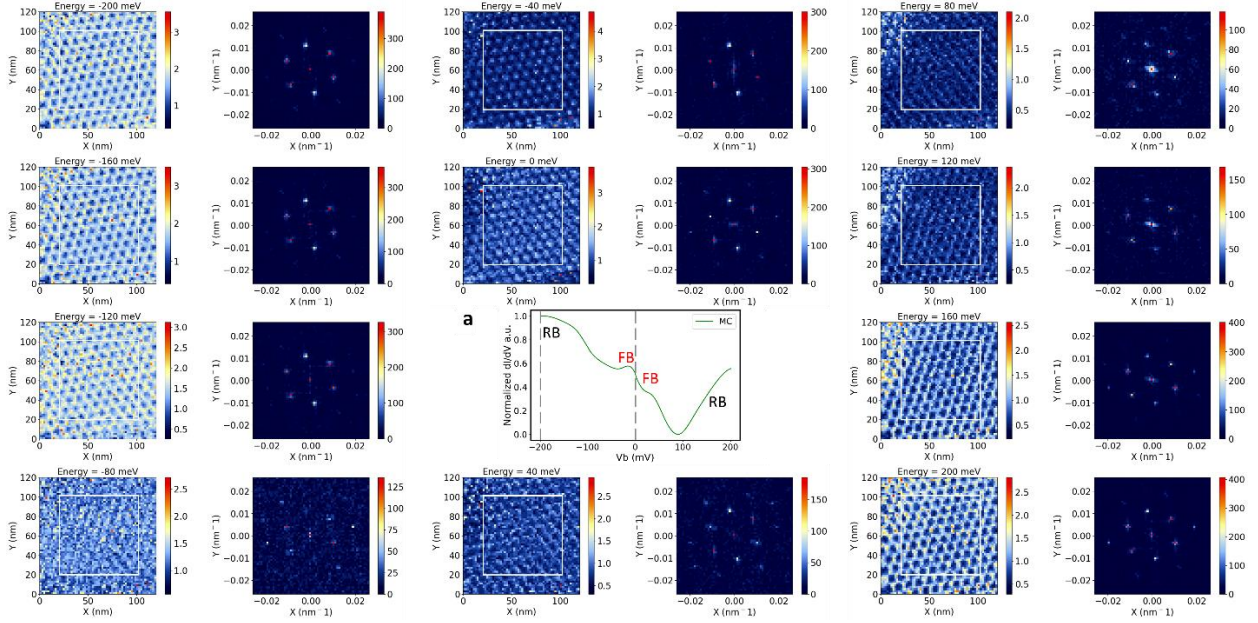

Fig. S18. **Additional LDoS mapping of moiré crystal.** The center panel (**a**) shows averaged dI/dV from the LDoS mapping of a moiré crystal region with $L_{GG} = L_{GBN} = 10.8\ nm$ in Fig. 54b. The moiré flat band are marked as FB and the remote bands are marked as RB. The rest of surrounding figures are LDoS mappings (left panels) and the corresponding FFTs (right panels) of the moiré crystal region presented in Fig. 5 at different $V_B$ (from -200 mV to 200 mV). The energies of the mappings are included in their titles. The FFT only shows peaks at the 1st and 2nd BZ, indicating the C3 symmetry of the real space pattern. We note that there is a domain boundary at the upper left corner of the mapping region. Fig. 5b is from the zoom-in to the regions in the white square. Tunneling parameters: $V_B$ = -300mV; $V_G$=10V; I = 100pA

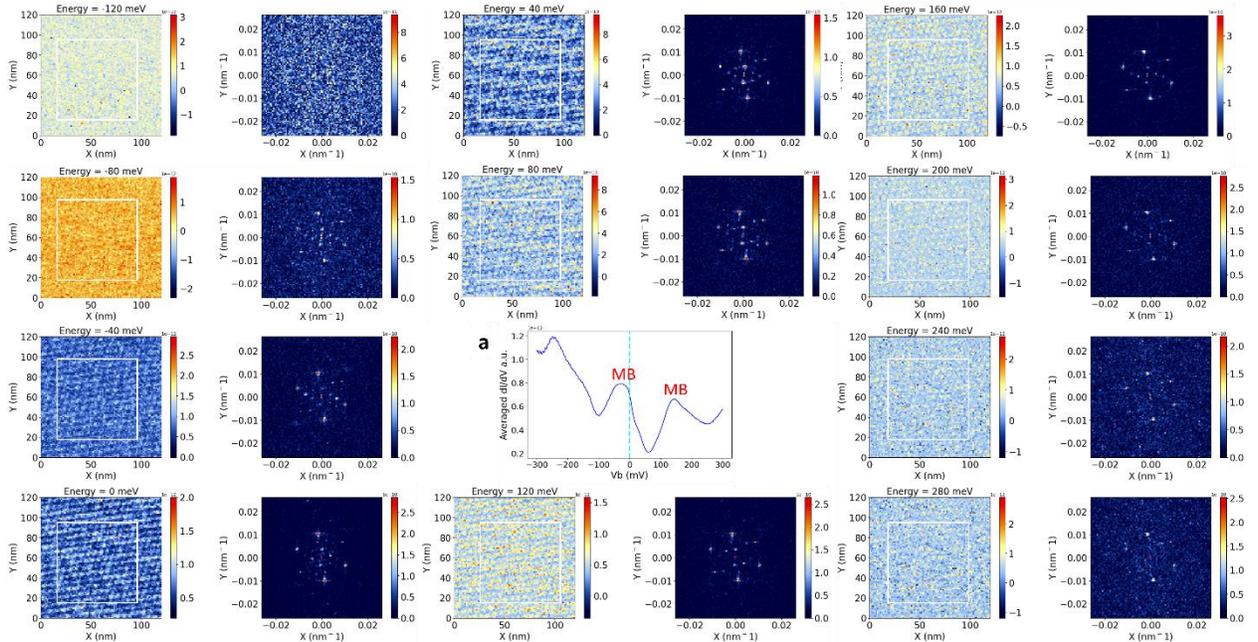

Fig. S19. **Additional LDoS Mapping of MIC.** The center panel (**a**) is averaged dI/dV from the LDoS mapping of the MIC region in Fig. 5f with $L_{GG} = 7.1 nm; L_{GBN} = 10.9\ nm$. The moiré bands are marked as MB in the center panel, the averaged dI/dV from all points in the map. The rest of surrounding figures are LDoS mappings (left panels) and the corresponding FFTs (right panels) of the MIC presented in Fig. 5f region at different $V_B$ (from -200

mV to 200 mV). The energies of the mappings are included in their titles. Fig. 5f is from the zoom-in to the regions in the white square. Tunneling parameters: $V_B$ = -300mV; $V_G$=0V; I = 100pA

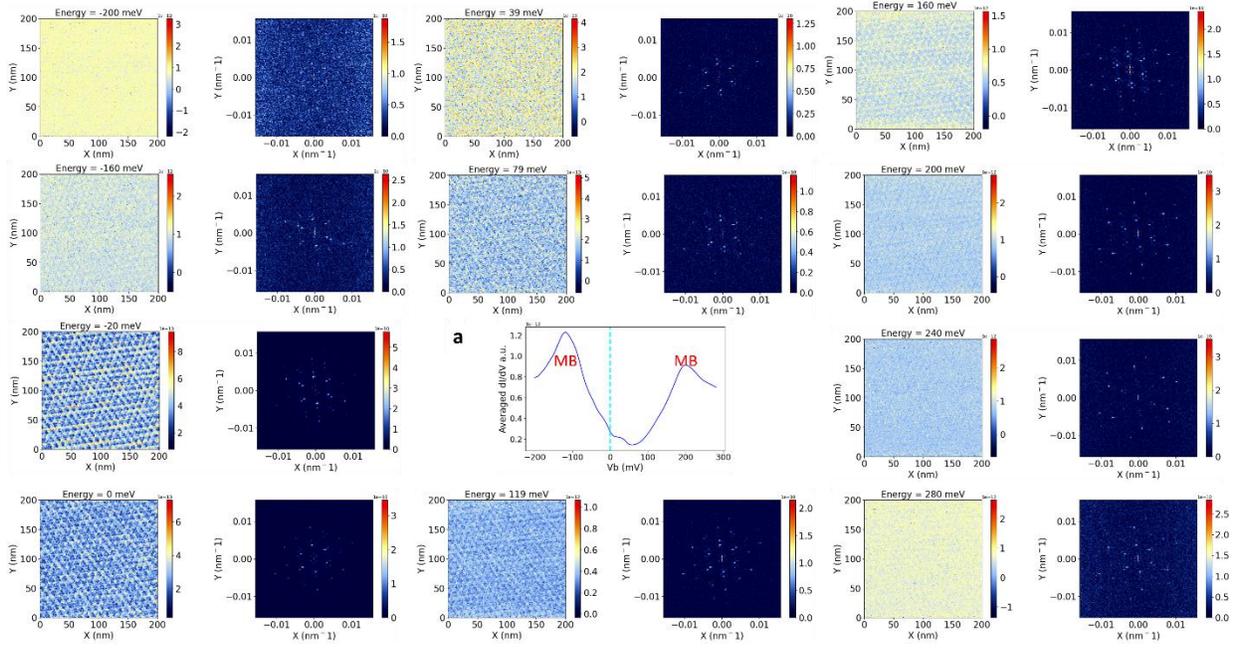

Fig. S20 **Additional LDoS mapping of a MIC with the directions of $\vec{K}_{GG}$ and $\vec{K}_{GBN}$ aligned.** The center panel (**a**) is averaged dI/dV from the LDoS mapping of the MIC region in Fig. S17 with $L_{GG} = 5.43\ nm; L_{GBN} = 12.47\ nm$. The moiré bands are marked as MB in the center panel, the averaged dI/dV from all points in the map. The rest of surrounding figures are LDoS mappings (left panels) and the corresponding FFTs (right panels) of the MIC region at different Vb (from -200 mV to 200 mV). The energies of the mappings are included in their titles. Tunneling parameters: $V_B$ = -200mV; $V_G$=0V; I = 100pA

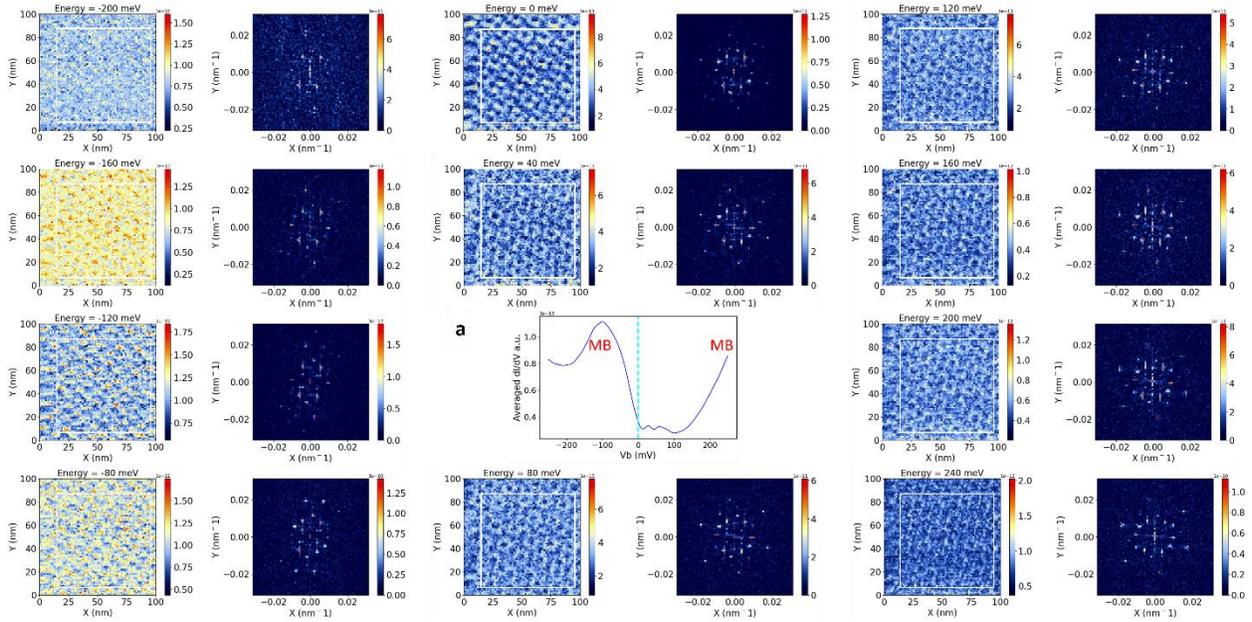

Fig. S21. **Additional LDoS Mapping of MQC.** The center panel (**a**) is averaged dI/dV from the LDoS mapping of the MQC region in Fig. S4j. The rest of surrounding figures are LDoS mappings (left panels) and the corresponding FFTs (right panels) of the MQC region at different Vb (from -200 mV to 240 mV) indicated by their titles. The intensity of GG, GBN and MM periods has different energy dependence, giving rise to the real space pattern with evolving symmetries. Fig. 5j is from the zoom-in to the regions in the white square. Tunneling parameters: $V_B$ = -300mV; $V_G$=0V; I = 100pA